\PassOptionsToPackage{unicode}{hyperref}
\PassOptionsToPackage{hyphens}{url}
\PassOptionsToPackage{dvipsnames,svgnames,x11names}{xcolor}
\documentclass[
  11pt,
  a4paper,
]{article}
\usepackage{xcolor}
\usepackage[left=2.5cm,right=2.5cm,top=2.5cm,bottom=3cm]{geometry}
\usepackage{amsmath,amssymb}
\usepackage{iftex}
\ifPDFTeX
  \usepackage[T1]{fontenc}
  \usepackage[utf8]{inputenc}
  \usepackage{textcomp} % provide euro and other symbols
\else % if luatex or xetex
  \ifXeTeX
    \usepackage{mathspec} % this also loads fontspec
  \else
    \usepackage{unicode-math} % this also loads fontspec
  \fi
  \defaultfontfeatures{Scale=MatchLowercase}
  \defaultfontfeatures[\rmfamily]{Ligatures=TeX,Scale=1}
\fi
\usepackage{lmodern}
\ifPDFTeX\else
\fi
\IfFileExists{upquote.sty}{\usepackage{upquote}}{}
\IfFileExists{microtype.sty}{% use microtype if available
  \usepackage[expansion=false]{microtype}
  \UseMicrotypeSet[protrusion]{basicmath} % disable protrusion for tt fonts
}{}
\makeatletter
\ifx\paragraph\undefined\else
  \let\oldparagraph\paragraph
  \renewcommand{\paragraph}{
    \@ifstar
      \xxxParagraphStar
      \xxxParagraphNoStar
  }
  \newcommand{\xxxParagraphStar}[1]{\oldparagraph*{#1}\mbox{}}
  \newcommand{\xxxParagraphNoStar}[1]{\oldparagraph{#1}\mbox{}}
\fi
\ifx\subparagraph\undefined\else
  \let\oldsubparagraph\subparagraph
  \renewcommand{\subparagraph}{
    \@ifstar
      \xxxSubParagraphStar
      \xxxSubParagraphNoStar
  }
  \newcommand{\xxxSubParagraphStar}[1]{\oldsubparagraph*{#1}\mbox{}}
  \newcommand{\xxxSubParagraphNoStar}[1]{\oldsubparagraph{#1}\mbox{}}
\fi
\makeatother

\usepackage{color}
\usepackage{fancyvrb}

\DefineVerbatimEnvironment{Highlighting}{Verbatim}{commandchars=\\\{\}}
\usepackage{framed}
\definecolor{shadecolor}{RGB}{244,242,239}
\newenvironment{Shaded}{\begin{snugshade}}{\end{snugshade}}

\newcommand{\AttributeTok}[1]{\textcolor[rgb]{0.35,0.33,0.01}{#1}}

\newcommand{\CommentTok}[1]{\textcolor[rgb]{0.55,0.49,0.45}{\textit{#1}}}

\newcommand{\ConstantTok}[1]{\textcolor[rgb]{0.50,0.29,0.00}{#1}}

\newcommand{\DecValTok}[1]{\textcolor[rgb]{0.50,0.29,0.00}{#1}}

\newcommand{\ErrorTok}[1]{\textcolor[rgb]{0.61,0.11,0.14}{\textbf{#1}}}

\newcommand{\FloatTok}[1]{\textcolor[rgb]{0.50,0.29,0.00}{#1}}
\newcommand{\FunctionTok}[1]{\textcolor[rgb]{0.43,0.14,0.39}{#1}}

\newcommand{\NormalTok}[1]{\textcolor[rgb]{0.07,0.08,0.09}{#1}}

\newcommand{\OtherTok}[1]{\textcolor[rgb]{0.00,0.23,0.46}{#1}}

\newcommand{\SpecialCharTok}[1]{\textcolor[rgb]{0.33,0.33,0.33}{#1}}

\newcommand{\StringTok}[1]{\textcolor[rgb]{0.00,0.30,0.35}{#1}}

\usepackage{longtable,booktabs,array}
\usepackage{calc} % for calculating minipage widths
\usepackage{etoolbox}
\makeatletter
\patchcmd\longtable{\par}{\if@noskipsec\mbox{}\fi\par}{}{}
\makeatother
\IfFileExists{footnotehyper.sty}{\usepackage{footnotehyper}}{\usepackage{footnote}}
\makesavenoteenv{longtable}
\usepackage{graphicx}
\makeatletter
\newsavebox\pandoc@box
\newcommand*\pandocbounded[1]{% scales image to fit in text height/width
  \sbox\pandoc@box{#1}%
  \Gscale@div\@tempa{\textheight}{\dimexpr\ht\pandoc@box+\dp\pandoc@box\relax}%
  \Gscale@div\@tempb{\linewidth}{\wd\pandoc@box}%
  \ifdim\@tempb\p@<\@tempa\p@\let\@tempa\@tempb\fi% select the smaller of both
  \ifdim\@tempa\p@<\p@\scalebox{\@tempa}{\usebox\pandoc@box}%
  \else\usebox{\pandoc@box}%
  \fi%
}
\def\fps@figure{htbp}
\makeatother

\providecommand{\tightlist}{%
  \setlength{\itemsep}{0pt}\setlength{\parskip}{0pt}}

\usepackage[style=apa,backend=biber,sorting=nyt,doi=true,url=true,isbn=false,uniquename=false,uniquelist=false]{biblatex}
\usepackage{booktabs}
\usepackage{caption}
\usepackage{longtable}
\usepackage{colortbl}
\usepackage{array}
\usepackage{anyfontsize}
\usepackage{multirow}
\DeclareFieldFormat[article]{volume}{\textit{\apanum{#1}\nocorr}}
\DeclareFieldFormat[inproceedings,periodical]{volume}{\textit{\apanum{#1}\nocorr}}
\usepackage{setspace}
\newcommand{\N}{\mathrm{N}}
\DeclareMathOperator*{\argmax}{arg\,max}

\usepackage{newunicodechar}
\usepackage{pifont}
\usepackage{textgreek}
\newlength{\consolesymwd}
\newcommand{\consolesym}[1]{\settowidth{\consolesymwd}{\textnormal{\checkmark}}\makebox[\consolesymwd][c]{\textnormal{#1}}}
\newunicodechar{✔}{\consolesym{\checkmark}}
\newunicodechar{ℹ}{\consolesym{\textit{i}}}
\newunicodechar{δ}{\ensuremath{\delta}}
\newunicodechar{σ}{\ensuremath{\sigma}}
\AtBeginDocument{%
  \definecolor{kblu}{HTML}{003B75}%
  \definecolor{korg}{HTML}{804900}%
  \definecolor{ktur}{HTML}{004C59}%
  \definecolor{ktursat}{HTML}{00A6AA}% KAUST saturated turquoise
  \definecolor{kturlink}{HTML}{3c8277}
  \definecolor{quarto-callout-note-color}{HTML}{5284C4}%
  \definecolor{quarto-callout-note-color-frame}{HTML}{5284C4}%
  \hypersetup{citecolor=kturlink,linkcolor=kturlink,urlcolor=kturlink}%
}
\makeatletter
\AtBeginDocument{%
  \@ifundefined{c@lemma}{}{%
    \@ifundefined{c@corollary}{}{%
      \let\c@corollary\c@lemma
    }%
    \@ifundefined{c@proposition}{}{%
      \let\c@proposition\c@lemma
    }%
  }%
}
\makeatother
\makeatletter
\@ifpackageloaded{tcolorbox}{}{\usepackage[skins,breakable]{tcolorbox}}
\@ifpackageloaded{fontawesome5}{}{\usepackage{fontawesome5}}
\definecolor{quarto-callout-color}{HTML}{909090}
\definecolor{quarto-callout-note-color}{HTML}{0758E5}
\definecolor{quarto-callout-important-color}{HTML}{CC1914}
\definecolor{quarto-callout-warning-color}{HTML}{EB9113}
\definecolor{quarto-callout-tip-color}{HTML}{00A047}
\definecolor{quarto-callout-caution-color}{HTML}{FC5300}
\definecolor{quarto-callout-color-frame}{HTML}{acacac}
\definecolor{quarto-callout-note-color-frame}{HTML}{4582ec}
\definecolor{quarto-callout-important-color-frame}{HTML}{d9534f}
\definecolor{quarto-callout-warning-color-frame}{HTML}{f0ad4e}
\definecolor{quarto-callout-tip-color-frame}{HTML}{02b875}
\definecolor{quarto-callout-caution-color-frame}{HTML}{fd7e14}
\makeatother
\makeatletter
\@ifpackageloaded{caption}{}{\usepackage{caption}}
\AtBeginDocument{%
\ifdefined\contentsname
  \renewcommand*\contentsname{Table of contents}
\else
  \newcommand\contentsname{Table of contents}
\fi
\ifdefined\listfigurename
  \renewcommand*\listfigurename{List of Figures}
\else
  \newcommand\listfigurename{List of Figures}
\fi
\ifdefined\listtablename
  \renewcommand*\listtablename{List of Tables}
\else
  \newcommand\listtablename{List of Tables}
\fi
\ifdefined\figurename
  \renewcommand*\figurename{Figure}
\else
  \newcommand\figurename{Figure}
\fi
\ifdefined\tablename
  \renewcommand*\tablename{Table}
\else
  \newcommand\tablename{Table}
\fi
}
\@ifpackageloaded{float}{}{\usepackage{float}}
\floatstyle{ruled}
\@ifundefined{c@chapter}{\newfloat{codelisting}{h}{lop}}{\newfloat{codelisting}{h}{lop}[chapter]}
\floatname{codelisting}{Listing}

\usepackage{amsthm}
\theoremstyle{remark}
\AtBeginDocument{}

\newtheorem{refremark}{Remark}[section]

\makeatother
\makeatletter
\@ifpackageloaded{caption}{}{\usepackage{caption}}
\@ifpackageloaded{subcaption}{}{\usepackage{subcaption}}
\makeatother
\usepackage{bookmark}
\IfFileExists{xurl.sty}{\usepackage{xurl}}{} % add URL line breaks if available
\hypersetup{
  pdftitle={Implementation and Workflows for INLA-Based Approximate Bayesian Structural Equation Modelling},
  pdfkeywords={Bayesian Structural Equation Modelling, Approximate Bayesian Inference, Integrated Nested Laplace Approximation (INLA), R Software, Computational Psychometrics},
  colorlinks=true,
  linkcolor={blue},
  filecolor={Maroon},
  citecolor={Blue},
  urlcolor={Blue},
  pdfcreator={LaTeX via pandoc}}

\title{Implementation and Workflows for INLA-Based Approximate Bayesian Structural Equation Modelling}
\usepackage{authblk}

\IfFileExists{orcidlink.sty}{\usepackage{orcidlink}}{\providecommand{\orcidlink}[1]{}}
\author[1,2]{Haziq Jamil~\orcidlink{0000-0003-3298-1010}\thanks{Correspondence concerning this article should be addressed to Haziq Jamil, Computer, Electrical and Mathematical Sciences and Engineering (CEMSE) Division, King Abdullah University of Science and Technology, Thuwal, 23955-6900, Kingdom of Saudi Arabia. E-mail: \href{mailto:haziq.jamil@kaust.edu.sa}{haziq.jamil@kaust.edu.sa}.}}
\author[1]{Håvard Rue~\orcidlink{0000-0002-0222-1881}}
\affil[1]{Computer, Electrical and Mathematical Sciences and Engineering (CEMSE) Division, King Abdullah University of Science and Technology}
\affil[2]{Mathematical Sciences, Faculty of Science, Universiti Brunei Darussalam}
\date{October 29, 2026}
\begin{document}
\maketitle
\begin{abstract}
Bayesian structural equation modelling (BSEM) offers many advantages such as principled uncertainty quantification, small-sample regularisation, and flexible model specification. However, the Markov chain Monte Carlo (MCMC) methods on which it relies can be slow for the iterative cycle of specification, criticism, and refinement that careful psychometric practice demands. We present \texttt{INLAvaan}, an R package for fast, approximate Bayesian SEM built around the Integrated Nested Laplace Approximation (INLA) framework for structural equation models developed by Jamil and Rue (2026, \href{https://doi.org/10.48550/arXiv.2603.25690}{\texttt{arXiv:2603.25690\ {[}stat.ME{]}}}). This paper serves as a companion manuscript that describes the architectural decisions and computational strategies underlying the package. Two substantive applications---a 256-parameter bifactor circumplex model and a multilevel mediation model with full-information missing-data handling---demonstrate how \texttt{INLAvaan} makes the iterative Bayesian workflow feasible in applied psychometric work.

\end{abstract}

\section{Introduction}\label{introduction}

Structural equation models (SEMs) remain the bedrock of modern psychometrics, offering a principled framework for inferring unobserved constructs from noisy measurements \autocite{hoyle2023handbook}.
While frequentist estimators such as maximum likelihood (ML) have long dominated the field due to their computational efficiency, the last two decades have seen a decisive shift towards Bayesian SEM (BSEM) \autocite{scheines1999bayesian,lee2007structural,song2012basic,muthen2012bayesian,maccallum2012hopes,palomo2007bayesian}.
Advances in computing made Markov chain Monte Carlo (MCMC) sampling practical, and user-friendly software then brought Bayesian SEM within reach of applied researchers.
The Bayesian approach also addresses at least three recurring problems of classical ML for SEM:

\begin{itemize}
\item
  \textbf{Inadmissible solutions.} ML is well known to produce Heywood cases (negative residual variances or out-of-range correlations), particularly in complex or small-sample models. Bayesian priors provide a natural regularisation that keeps estimates within admissible bounds \autocite{rindskopf1984structural,chen2001improper,martin1975bayesian}.
\item
  \textbf{Small-sample breakdown.} In small-sample settings, frequentist methods such as ML suffer from poor inferential performance because the necessary asymptotic assumptions are not satisfied \autocite{jamil2026biasreduced}. Bayesian inference with appropriately chosen priors absorbs this uncertainty more reliably, reducing bias and improving coverage \autocite{lee2004evaluation,mcneish2016usinga,ulitzsch2023alleviating,stegmueller2013how,smid2020bayesian}.
\item
  \textbf{Flexible model specification.} Many theoretically motivated models require constraints that classical point-null hypothesis testing handles poorly, such as approximate zero constraints (cross-loadings assigned a small-variance prior rather than fixed to zero), inequality restrictions to enforce sign patterns, or informative priors to resolve the scale and rotational indeterminacies inherent to latent variable models \autocite{muthen2012bayesian,vandeschoot2011moving,erosheva2017dealing,vanerp2023bayesian}.
\end{itemize}

A rich ecosystem of software for Bayesian SEM now exists, ranging from ``ready-to-use'' platforms to flexible probabilistic programming languages.
On the proprietary side, \texttt{Mplus} \autocite{muthen2012bayesian,muthen2017mplus}, IBM's \texttt{AMOS} \autocite{arbuckle2022ibm}, and \texttt{Stata} \autocite{statacorp2025stata} are the dominant commercial platforms.
These packages support a broad range of normal-theory SEM classes (e.g., multigroup, multilevel, and growth curve models) with well-documented default priors, posterior predictive model checking, and accessible interfaces that have made them the de facto standards in applied social science, though their closed-source nature limits reproducibility and community-driven extensions.

Within the open-source R ecosystem, \texttt{blavaan} \autocite{merkle2018blavaan,merkle2021efficient} provides a seamless Bayesian extension of \texttt{lavaan} \autocite{rosseel2012lavaan}, automatically translating familiar \texttt{lavaan} syntax into \texttt{Stan} or \texttt{JAGS} sampling code.
Its feature set has matured substantially, making it a genuine open-source alternative to \texttt{Mplus} for a wide range of normal-theory SEMs.
Indicative of its adoption, \texttt{blavaan} records approximately 8,200 downloads per month on CRAN (2026 average), building on the large user base of \texttt{lavaan}.
\texttt{brms} \autocite{burkner2021bayesian} offers a broader multilevel modelling interface via \texttt{Stan} for researchers willing to step outside the \texttt{lavaan} syntax convention.
Python users are served by \texttt{semba} \autocite{meshcheryakov2021semopy}, a \texttt{NumPyro}-backed Bayesian counterpart to the frequentist \texttt{semopy} library \autocite{igolkina2020semopy}.

For researchers requiring fully bespoke model specification, general-purpose probabilistic programming languages can encode any normal-theory SEM directly, at the cost of hand-coding the model in its entirety without providing any psychometric convenience layer.
This not only includes the aforementioned \texttt{Stan} \autocite{carpenter2017stan}, \texttt{JAGS} \autocite{plummer2003jags}, and \texttt{NumPyro} \autocite{phan2019composable,bingham2019pyro}, but also \texttt{PyMC} \autocite{pymc2023}.

Despite their variety, all of these tools share the same computational engine, MCMC sampling, which can be slow by the standards of iterative applied work and creates a significant workflow gap in practice.
Psychometric analysis in particular is rarely a single-shot estimation problem; it is an iterative cycle of specification, criticism, and refinement \autocite{maccallum1986specification,bollen1993testing}, as the Bayesian workflow requires \autocite{gelman2026bayesian}.
A typical measurement invariance study, for example, requires fitting a sequence of configural, metric, and scalar models across groups, diagnosing misfit at each step, and revising the specification accordingly \autocite{meredith1993measurement}.
Each of these fits can demand significant MCMC run time.
Together with convergence diagnostics, chain inspection, and potential restarts, this makes the feedback loop slow.
The result is a difficult compromise.
Researchers who value rapid, exploratory model building are effectively pushed back towards frequentist estimators, forfeiting the inferential benefits of the Bayesian framework in the complex settings where those benefits matter most.

To avoid this compromise, we introduce \texttt{INLAvaan}, an R package that provides fast, approximate Bayesian SEM through a \emph{ground-up implementation} of the Integrated Nested Laplace Approximation \autocite[INLA;][]{rue2009approximate} that is \textbf{entirely independent of the general-purpose \texttt{R-INLA} software} and engineered specifically for the \texttt{lavaan} modelling framework \autocite{rosseel2012lavaan}.
\texttt{INLAvaan} inherits \texttt{lavaan}'s syntax, parameter table, and model algebra, but replaces the fitting engine with a deterministic approximation of the SEM posterior distributions.
For normal-theory SEM, the likelihood of the observed variables, marginal over the latent variables, is available in closed form, so the posterior involves only the model parameters, typically numbering a few dozen to a few hundred.
This posterior is then obtained by optimisation and Laplace-type expansions instead of sampling, with no chains to run or monitor.
This approach substantially reduces the computational cost of iteration in the Bayesian workflow, allowing researchers to explore and refine model specifications with the rigour that careful psychometric practice demands.

This manuscript serves as the computational counterpart to the theoretical foundations in \textcite{jamil2026approximate}.
Where that work establishes the statistical methodology and formal proofs of accuracy, this paper focuses on the architectural decisions and computational implementation required to make these advances accessible in practice.
Accordingly, our primary objective here is to demonstrate \textbf{workflow efficiency, computational scalability, and applied modelling possibilities}.
Each stage of the pipeline is detailed, including analytic Jacobian assembly, variational location correction, skew-normal marginal fitting, and Gaussian copula reconstruction, showing how each is realised as fast R computations.
Two substantive use cases, a circumplex factor model that also runs one pass of the Bayesian workflow (Section \ref{sec-circfa}) and a multilevel safety-climate analysis (Section \ref{sec-safety}), illustrate the practical benefits of the approach on substantive psychometric models.

\section{\texorpdfstring{Bayesian SEM using \texttt{INLAvaan}}{Bayesian SEM using INLAvaan}}\label{sec-bsem}

This section establishes the model and prior specification underlying \texttt{INLAvaan}, walks through a complete worked example, and describes the range of SEM variants the package supports.

\subsection{Preliminaries}\label{sec-preliminaries}

As SEM is well established and treated extensively in the literature \autocites[e.g.,][]{bollen1989structural,kaplan2009structural,hoyle2023handbook}, we cover its specification only briefly here, primarily to fix notation.
Let \(\mathbf{y}_s \in \mathbb{R}^p\) denote the observed response vector for subject \(s = 1, \dots, n\).
The normal-theory SEM comprises a measurement equation linking observed indicators to latent factors \(\boldsymbol{\eta}_s \in \mathbb{R}^q\), \(q \ll p\), and a structural equation governing the relations among those factors:
\begin{equation}\protect\phantomsection\label{eq-sem-equations}{
\begin{aligned}
\mathbf{y}_s &= \boldsymbol{\nu} + \boldsymbol{\Lambda} \boldsymbol{\eta}_s + \boldsymbol{\epsilon}_s, \quad \boldsymbol{\epsilon}_s \sim \N_p(\mathbf{0}, \boldsymbol{\Theta}), \\
\boldsymbol{\eta}_s &= \boldsymbol{\alpha} + \mathbf{B} \boldsymbol{\eta}_s + \boldsymbol{\zeta}_s, \quad \boldsymbol{\zeta}_s \sim \N_q(\mathbf{0}, \boldsymbol{\Psi}).
\end{aligned}
}\end{equation}
Here, \(\boldsymbol{\nu}\) is a \(p\)-vector of measurement intercepts, \(\boldsymbol{\Lambda}\) is a \(p \times q\) factor loading matrix, \(\boldsymbol{\Theta}\) is the measurement error covariance, \(\boldsymbol{\alpha}\) is a \(q\)-vector of latent intercepts, \(\mathbf{B}\) is a \(q \times q\) matrix of structural regression coefficients, and \(\boldsymbol{\Psi}\) is the structural disturbance covariance.
We assume \(\boldsymbol{\epsilon}_s\) and \(\boldsymbol{\zeta}_s\) are independent, and that \((\mathbf{I} - \mathbf{B})\) is invertible.

We collect the unique, estimable free entries of
\(\{\boldsymbol{\nu}, \boldsymbol{\Lambda}, \boldsymbol{\Theta}, \boldsymbol{\alpha}, \mathbf{B}, \boldsymbol{\Psi}\}\)
into the \emph{natural parameter vector} \(\mathbf{x} \in \mathbb{R}^m\), i.e., only the non-redundant elements that are not fixed by identification constraints (e.g., scale-setting or exclusion restrictions) and that are not deterministic functions of other parameters.
Integrating out the latent variables yields the marginal distribution \(\mathbf{y}_s \mid \mathbf{x} \sim \N_p\big(\boldsymbol{\mu}(\mathbf{x}),\, \boldsymbol{\Sigma}(\mathbf{x})\big)\), with implied moments
\begin{equation}\protect\phantomsection\label{eq-implied-moments}{
\begin{aligned}
\boldsymbol{\mu}(\mathbf{x}) &= \boldsymbol{\nu} + \boldsymbol{\Lambda}(\mathbf{I} - \mathbf{B})^{-1}\boldsymbol{\alpha}, \\
\boldsymbol{\Sigma}(\mathbf{x}) &= \boldsymbol{\Lambda}(\mathbf{I} - \mathbf{B})^{-1}\boldsymbol{\Psi}(\mathbf{I} - \mathbf{B})^{-\top}\boldsymbol{\Lambda}^\top + \boldsymbol{\Theta}.
\end{aligned}
}\end{equation}
Bayesian inference targets the posterior \(\pi(\mathbf{x} \mid \mathbf{y})\) by combining a prior \(\pi(\mathbf{x})\) with the Gaussian log-likelihood evaluated efficiently via the sufficient statistics \(\bar{\mathbf{y}} = n^{-1} \sum_{s=1}^n \mathbf{y}_s\) and \(\mathbf{S} = n^{-1} \sum_{s=1}^n (\mathbf{y}_s - \bar{\mathbf{y}})(\mathbf{y}_s - \bar{\mathbf{y}})^\top\).
Where MCMC would draw samples from this posterior, \texttt{INLAvaan} approximates it deterministically (Section~\ref{sec-implementation}).

We assign each type of parameter a prior family suited to its range.
Intercepts, loadings, and regression coefficients (\(\boldsymbol{\nu}\), \(\boldsymbol{\alpha}\), \(\boldsymbol{\Lambda}\), \(\mathbf{B}\)) take normal priors.
For the covariance matrices, we adopt the separation strategy of \textcite{barnard2000modeling}, as in \textcite{merkle2021efficient}, where each matrix (e.g., \(\boldsymbol{\Psi}\)) is decomposed as \(\boldsymbol{\Psi} = \mathbf{D}_\psi\, \mathbf{R}_\psi\, \mathbf{D}_\psi\), where \(\mathbf{D}_\psi = \operatorname{diag}(\Psi_{11}^{1/2}, \dots, \Psi_{qq}^{1/2})\) holds the standard deviations and \(\mathbf{R}_\psi\) is a correlation matrix.
The standard deviations then take gamma priors, and each off-diagonal entry of \(\mathbf{R}_\theta\) and \(\mathbf{R}_\psi\) takes a shifted beta prior on \((-1,1)\).
This places priors directly on interpretable quantities (scales and correlations), although separate priors on the correlations do not by themselves guarantee positive definiteness of the reconstructed matrix (see Remark~\ref{rem-pd-blocks}).
The default hyperparameters are listed in Table 1 of \textcite{jamil2026approximate} and appear in the \texttt{Prior} column of the \texttt{summary()} output in Section~\ref{sec-worked-example}.

When the means of the observed variables are left unstructured (\texttt{meanstructure\ =\ FALSE}, the \texttt{lavaan} default, and also ours, for complete single-group data), \texttt{INLAvaan} does not profile them out at \(\bar{\mathbf y}\) as \texttt{lavaan} does, but gives them an improper uniform prior and integrates them out from the likelihood, which yields
\[
\int_{\mathbb R^p} \prod_{s=1}^n \phi_p(\mathbf y_s;\, \boldsymbol\mu, \boldsymbol\Sigma)\, d\boldsymbol\mu
\;\propto\; |\boldsymbol\Sigma|^{-(n-1)/2} \exp\Big\{-\tfrac{n-1}{2}\operatorname{tr}\big(\boldsymbol\Sigma^{-1}\mathbf S^*\big)\Big\},
\]
where \(\phi_p\) is the \(p\)-variate normal density and \(\mathbf S^* = n\mathbf S/(n-1)\).
This is the Wishart likelihood for \(\boldsymbol\Sigma\), based on \(\mathbf S^*\) with \(n-1\) degrees of freedom.

\subsection{A Worked Example}\label{sec-worked-example}

Given the model and prior specification above, \texttt{INLAvaan} approximates \(\pi(\mathbf{x} \mid \mathbf{y})\) through a sequence of deterministic steps developed in \textcite{jamil2026approximate} and depicted schematically in Figure~\ref{fig-inlavaan-recipe}.
We illustrate the pipeline using Bollen's \autocite*{bollen1989structural} oft-cited three-factor SEM in which industrialisation in 1960 (\texttt{ind60}) predicts political democracy in 1960 (\texttt{dem60}) and 1965 (\texttt{dem65}), with six residual covariances among repeated indicators, giving \(m = 31\) free parameters without a mean structure.
The model is set up as a string vector, like so:

\begin{Shaded}
\begin{Highlighting}[]
\NormalTok{mod\_poldem }\OtherTok{\textless{}{-}} \StringTok{"}
\StringTok{  \# Latent variable definitions}
\StringTok{    ind60 =\textasciitilde{} x1 + x2 + x3}
\StringTok{    dem60 =\textasciitilde{} y1 + y2 + y3 + y4}
\StringTok{    dem65 =\textasciitilde{} y5 + y6 + y7 + y8}
\StringTok{  \# Latent regressions}
\StringTok{    dem60 \textasciitilde{} ind60}
\StringTok{    dem65 \textasciitilde{} ind60 + dem60}
\StringTok{  \# Residual correlations}
\StringTok{    y1 \textasciitilde{}\textasciitilde{} y5}
\StringTok{    y2 \textasciitilde{}\textasciitilde{} y4 + y6}
\StringTok{    y3 \textasciitilde{}\textasciitilde{} y7}
\StringTok{    y4 \textasciitilde{}\textasciitilde{} y8}
\StringTok{    y6 \textasciitilde{}\textasciitilde{} y8}
\StringTok{"}
\end{Highlighting}
\end{Shaded}

To fit the model, a single call to \texttt{asem()} runs the full approximation regime, with progress reported to the console.
As a mirror interface to \texttt{lavaan::sem()}, \texttt{asem()} accepts the same model string and data frame, with the `\texttt{a}' prefix signalling the \emph{approximate} Bayesian engine throughout the package.
The entire procedure completes in about 1.2 seconds on a MacBook Pro (Apple M4 Pro, 14-core CPU, 24 GB unified RAM), a turnaround close to that of a frequentist ML fit, with the priors and full posterior included.

\begin{Shaded}
\begin{Highlighting}[]
\NormalTok{R}\SpecialCharTok{\textgreater{}} \FunctionTok{library}\NormalTok{(INLAvaan)}
\NormalTok{R}\SpecialCharTok{\textgreater{}}\NormalTok{ fit }\OtherTok{\textless{}{-}} \FunctionTok{asem}\NormalTok{(}\AttributeTok{model =}\NormalTok{ mod\_poldem, }\AttributeTok{data =}\NormalTok{ lavaan}\SpecialCharTok{::}\NormalTok{PoliticalDemocracy)}
\end{Highlighting}
\end{Shaded}

\begin{verbatim}
✔ Posterior mode and Hessian. [69ms]
✔ VB correction; mean |δ| = 0.208σ. [103ms]
✔ Fit 31/31 skew-normal marginals. [639ms]
✔ Summarise 1000 posterior draws. [315ms]
ℹ Fit measures: PPP, DIC.
\end{verbatim}

The console output reports each step of the approximation as it completes, and these steps fall into three stages.
All operate in an unconstrained parameterisation \(\boldsymbol{\vartheta}\), related to the natural parameters through element-wise bijections \(\{g_j\}\) and covariance reconstruction (Section~\ref{sec-param-space}).

\begin{enumerate}
\def\labelenumi{\arabic{enumi}.}
\tightlist
\item
  \textbf{Joint Gaussian approximation.} (a) A gradient-based optimiser locates the maximum a posteriori (MAP) estimate \(\boldsymbol{\vartheta}^*\) and evaluates the negative Hessian \(\mathbf{H}\) of the log-posterior at the mode, producing the Laplace approximation \(\N_m(\boldsymbol{\vartheta}^*,\, \mathbf{H}^{-1})\).
  (b) A VB step then shifts the Gaussian centre by \(\hat{\boldsymbol{\delta}}\) towards the posterior mean, correcting the joint location for the skewness of the posterior.
  Here the absolute shift averages 0.21 posterior standard deviations, which is small but not negligible (see Section~\ref{sec-vb}).
\item
  \textbf{Skew-normal marginals.} Each marginal \(\pi(\vartheta_j \mid \mathbf{y})\) is scanned along \(\vartheta_j\), with the other parameters set to their conditional means, corrected for the change in their spread along the scan (Section~\ref{sec-volume-correction}), and fitted with a skew-normal density.
  These marginals, not the joint Gaussian of Stage 1, give the posterior summaries.
\item
  \textbf{Copula sampling.} Joint samples are drawn from a Gaussian copula built on these marginals (Section~\ref{sec-copula}) to compute covariance parameters, fit indices, and other derived quantities.
\end{enumerate}

Calling \texttt{summary()} on the fit object returns posterior means, standard deviations, and credible intervals for the Political Democracy example, all reported on the natural \(\mathbf{x}\) scale.

\begin{Shaded}
\begin{Highlighting}[]
\NormalTok{R}\SpecialCharTok{\textgreater{}} \FunctionTok{summary}\NormalTok{(fit)}
\end{Highlighting}
\end{Shaded}

\begin{verbatim}
INLAvaan 0.3.3 ended normally after 65 iterations

  Estimator                                      BAYES
  Optimization method                           NLMINB
  Number of model parameters                        31

  Number of observations                            75

Model Test (User Model):

   Marginal log-likelihood                   -1649.518 
   PPP (Chi-square)                              0.519 

Information Criteria:

   Deviance (DIC)                             3173.920 
   Effective parameters (pD)                    30.139 

Parameter Estimates:

   Marginalisation method                     SKEWNORM
   VB correction                                  TRUE

Latent Variables:
                   Estimate       SD     2.5%    97.5%     NMAD    Prior       
  ind60 =~                                                                     
    x1                1.000                                                    
    x2                2.221    0.148    1.951    2.534    0.007    normal(0,10)
    x3                1.841    0.156    1.548    2.161    0.004    normal(0,10)
  dem60 =~                                                                     
    y1                1.000                                                    
    y2                1.312    0.196    0.954    1.722    0.005    normal(0,10)
    y3                1.092    0.153    0.808    1.410    0.004    normal(0,10)
    y4                1.327    0.166    1.033    1.683    0.009    normal(0,10)
  dem65 =~                                                                     
    y5                1.000                                                    
    y6                1.227    0.181    0.904    1.614    0.008    normal(0,10)
    y7                1.320    0.169    1.020    1.684    0.009    normal(0,10)
    y8                1.316    0.176    1.007    1.696    0.010    normal(0,10)

Regressions:
                   Estimate       SD     2.5%    97.5%     NMAD    Prior       
  dem60 ~                                                                      
    ind60             1.452    0.398    0.690    2.254    0.002    normal(0,10)
  dem65 ~                                                                      
    ind60             0.548    0.238    0.091    1.025    0.001    normal(0,10)
    dem60             0.845    0.100    0.662    1.053    0.010    normal(0,10)

Covariances:
                   Estimate       SD     2.5%    97.5%     NMAD    Prior       
 .y1 ~~                                                                        
   .y5                0.696    0.387   -0.005    1.514    0.002       beta(1,1)
 .y2 ~~                                                                        
   .y4                1.293    0.722   -0.021    2.814    0.005       beta(1,1)
   .y6                2.164    0.755    0.819    3.783    0.012       beta(1,1)
 .y3 ~~                                                                        
   .y7                0.905    0.645   -0.220    2.310    0.005       beta(1,1)
 .y4 ~~                                                                        
   .y8                0.346    0.477   -0.522    1.350    0.003       beta(1,1)
 .y6 ~~                                                                        
   .y8                1.364    0.574    0.376    2.623    0.005       beta(1,1)

Variances:
                   Estimate       SD     2.5%    97.5%     NMAD    Prior       
   .x1                0.090    0.022    0.053    0.138    0.006 gamma(1,.5)[sd]
   .x2                0.132    0.067    0.030    0.283    0.031 gamma(1,.5)[sd]
   .x3                0.509    0.101    0.341    0.735    0.003 gamma(1,.5)[sd]
   .y1                2.138    0.527    1.246    3.302    0.009 gamma(1,.5)[sd]
   .y2                7.805    1.455    5.354   11.034    0.002 gamma(1,.5)[sd]
   .y3                5.500    1.083    3.710    7.939    0.001 gamma(1,.5)[sd]
   .y4                3.320    0.811    1.917    5.085    0.013 gamma(1,.5)[sd]
   .y5                2.598    0.554    1.677    3.837    0.005 gamma(1,.5)[sd]
   .y6                5.264    0.974    3.622    7.425    0.002 gamma(1,.5)[sd]
   .y7                3.712    0.803    2.371    5.502    0.006 gamma(1,.5)[sd]
   .y8                3.443    0.754    2.154    5.103    0.006 gamma(1,.5)[sd]
    ind60             0.460    0.090    0.311    0.663    0.003 gamma(1,.5)[sd]
   .dem60             3.918    0.942    2.354    6.028    0.005 gamma(1,.5)[sd]
   .dem65             0.265    0.189    0.026    0.721    0.043 gamma(1,.5)[sd]
\end{verbatim}

The accuracy of the approximation is assessed in detail by \textcite[Sec. 4]{jamil2026approximate}, and we summarise their findings here.
For this benchmark model with a mean structure, the approximate marginals are practically indistinguishable from MCMC: Jensen--Shannon similarity is at least 98.5\% and standardised mean discrepancy at most 0.23 posterior standard deviations for every parameter.
Simulation-based calibration indicates a well-calibrated posterior under informative priors, with mild, localised departures under diffuse priors.
Simulations varying sample size, measurement quality, predictor collinearity, and model size show close agreement with MCMC when data are informative and parameters well identified.
Agreement declines gradually as identification weakens, consistent with the diagnostics of Section~\ref{sec-mode-hessian}.

\subsection{Generality of the Likelihood}\label{sec-likelihood-scope}

Although we presented (\ref{eq-sem-equations}) in its general SEM form, the same normal-theory likelihood already subsumes a wide family of models as special cases, namely confirmatory factor analysis \autocite[CFA;][Ch. 3]{bartholomew2011latent}, latent growth curves \autocite{skrondal2004generalized}, path models \autocite[Part II, Ch. 5]{kline2023principles}, mediation analyses \autocite{mackinnon2008introduction}, and even multiple linear regression \autocite[Ch. 6]{moustaki2025analysis}.
These all arise from specific restrictions on the matrices \(\boldsymbol{\Lambda}\), \(\mathbf{B}\), \(\boldsymbol{\Psi}\), and \(\boldsymbol{\Theta}\) in (\ref{eq-sem-equations}) and (\ref{eq-implied-moments}).
Following conventions in \texttt{lavaan} and \texttt{blavaan}, \texttt{INLAvaan} provides dedicated wrappers around \texttt{inlavaan()} for the most common model classes, namely \texttt{acfa()} for CFA, \texttt{agrowth()} for latent growth curves, and \texttt{asem()} for general SEM.

Beyond these canonical cases, many practically important extensions reduce to the same structural form, differing only in how data are partitioned and how parameters are shared across blocks.
In each case the log-likelihood is a sum of Gaussian log-densities over \(K\) blocks,
\begin{equation}\protect\phantomsection\label{eq-general-likelihood}{
\ell(\mathbf{x}) = \sum_{k=1}^K \log \phi_{p_k}\!\left(\mathbf{y}_k;\, \boldsymbol{\mu}_k(\mathbf{x}),\, \boldsymbol{\Sigma}_k(\mathbf{x})\right),
}\end{equation}
where \(k\) runs over whatever data blocks the model defines, \(\mathbf{y}_k \in \mathbb{R}^{p_k}\) collects the observations in block \(k\), and the implied moments \((\boldsymbol{\mu}_k, \boldsymbol{\Sigma}_k)\) follow the same algebra as (\ref{eq-implied-moments}) applied to each block.
Because the normal family is an exponential family, each block's contribution collapses via sufficiency to a kernel evaluated at the block-wise sample mean \(\bar{\mathbf{y}}_k\) and sample covariance \(\mathbf{S}_k\), and no per-observation loop is required.
The single-group complete-data case is just \(K = 1\) with \(p_1 = p\); the extensions below are instances of the same form:

\begin{itemize}
\item
  \textbf{Multigroup CFA/SEM.} \(k\) indexes \(G\) groups, each contributing sufficient statistics \((\bar{\mathbf{y}}_k, \mathbf{S}_k, n_k)\) and implied moments \((\boldsymbol{\mu}_k, \boldsymbol{\Sigma}_k)\). Parameters may be freely estimated per group or partially constrained equal; constraining loadings and then intercepts yields the configural \(\to\) metric \(\to\) scalar invariance hierarchy \autocite{meredith1993measurement}.
\item
  \textbf{Multilevel CFA/SEM.} \(k\) indexes \(J\) clusters, each containing \(n_k\) observations with \(p_k = n_k p\). Observations within a cluster are not independent, so the implied block covariance \(\boldsymbol{\Sigma}_k = \mathbf{I}_{n_k} \otimes \boldsymbol{\Sigma}_W + \mathbf{J}_{n_k} \otimes \boldsymbol{\Sigma}_B\) is a sum of Kronecker products, where \(\mathbf{J}_{n_k} = \mathbf{1}_{n_k}\mathbf{1}_{n_k}^\top\) encodes within-cluster exchangeability and \(\boldsymbol{\Sigma}_W\), \(\boldsymbol{\Sigma}_B\) are the within- and between-level implied covariance matrices shared across all \(J\) clusters. Exploiting the spectral structure of \(\mathbf{J}_{n_k}\), each cluster's likelihood reduces to two terms: a within-cluster scatter governed by \(\boldsymbol{\Sigma}_W\) and a cluster-mean term governed by \(\boldsymbol{\Sigma}_W + n_k\boldsymbol{\Sigma}_B\). These terms utilise the sufficient statistics \((\bar{\mathbf{y}}_k, \mathbf{S}_{W,k}, n_k)\), with all \(J\) blocks sharing the same \(\mathbf{x}\) \autocite{rosseel2021evaluating}.
\item
  \textbf{Missing data.} \(k\) indexes the \(R\) distinct missing-data patterns, and \(p_k\) is the number of observed items under pattern \(k\). The implied moments \((\boldsymbol{\mu}_k, \boldsymbol{\Sigma}_k)\) are simply the sub-vector and sub-matrix of the full implied moments restricted to those observed items, with all \(K = R\) blocks sharing the same \(\mathbf{x}\) \autocite[Ch. 5]{enders2022applied}.
\end{itemize}

As the approximation in \textcite{jamil2026approximate} is anchored entirely to this likelihood and its gradient, the entire recipe carries over to each case without modification, and only the implied-moments function \(\mathbf{x} \mapsto \{(\boldsymbol{\mu}_k, \boldsymbol{\Sigma}_k)\}\) is swapped.
The availability of closed-form gradients \autocite{neudecker1991linear,vonoertzen2014efficient} is especially important here:
with \(K\) blocks each requiring a \(p_k \times p_k\) matrix inversion, finite-difference approximation would scale poorly, whereas the analytic gradient keeps the per-iteration cost linear in \(K\) for every model class.

\section{Algorithmic Implementation}\label{sec-implementation}

\begin{figure}

\centering{

\includegraphics[width=1\linewidth,height=\textheight,keepaspectratio]{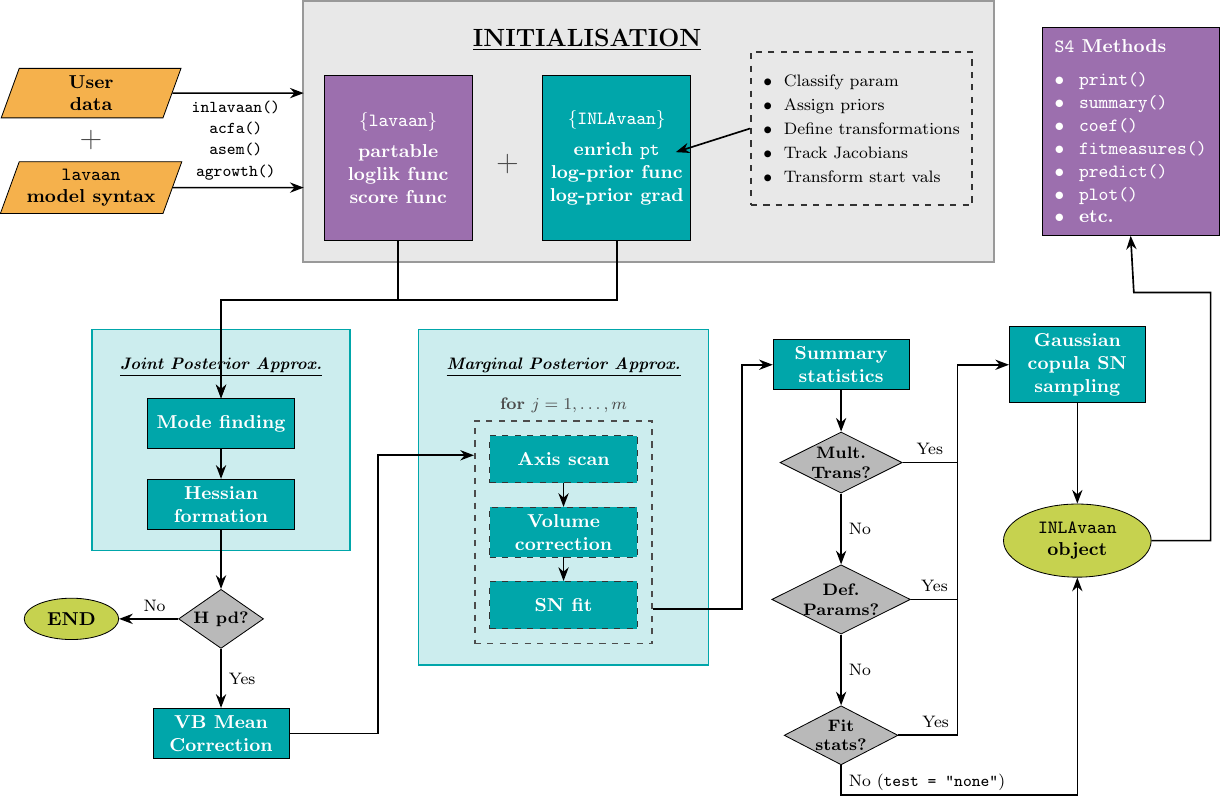}

}

\caption{\label{fig-inlavaan-recipe}The \texttt{INLAvaan} pipeline: after initialisation, the three stages of the approximation, (1) joint posterior approximation with the VB mean correction, (2) marginal profiling, and (3) Gaussian copula sampling, yield a fully fitted Bayesian SEM object.}

\end{figure}%

This section details the architectural decisions behind the INLA-based approximation of \textcite{jamil2026approximate}, depicted in Figure~\ref{fig-inlavaan-recipe}, focusing on the software engineering required to make them fast, correct, and diagnostic-rich within an R package built on top of \texttt{lavaan} \autocite{rosseel2012lavaan}.
Each subsection closes with a brief practical upshot, so the exposition will hopefully suit both technical and applied readers.

We first describe how the package wraps around \texttt{lavaan}.
The main user-facing function \texttt{inlavaan()} (and its wrappers \texttt{acfa()}, \texttt{asem()}, \texttt{agrowth()}) orchestrates the full pipeline in a single call.
To avoid reimplementing model algebra from scratch, \texttt{INLAvaan} seeds the pipeline with a zero-iteration \texttt{lavaan} object (\texttt{do.fit\ =\ FALSE}), inheriting syntax parsing, parameter table construction, model matrices, and sample statistics without modification, then overlays its own Bayesian machinery on top.
The result is an S4 object of class \texttt{INLAvaan} extending the \texttt{lavaan-class}, so every downstream \texttt{lavaan} method (\texttt{coef}, \texttt{summary}, \texttt{parameterestimates}, \texttt{fitmeasures}, etc.) continues to work unchanged.
This design means that adopting \texttt{INLAvaan} requires changing precisely one function call (e.g., \texttt{sem()} to \texttt{asem()}) and nothing else in the user's script, in order to get to a minimally working setup.

\subsection{Parameter Space and Constraint Handling}\label{sec-param-space}

The first implementation challenge is to convert a \texttt{lavaan} model specification, with its mix of loadings, variances, correlations, intercepts, and regression coefficients, into an unconstrained optimisation problem over \(\mathbb{R}^m\) suitable for gradient-based mode-finding and Laplace approximation.
Under the separation strategy adopted in Section~\ref{sec-bsem}, the working parameters include positivity-constrained quantities (variances on \((0,\infty)\)) and bounded ones (correlations on \((-1,1)\)).
An optimiser tasked with navigating these constrained regions can encounter boundary singularities and ill-conditioned Hessians, both common sources of convergence failure in applied Bayesian SEM.

Good starting values may keep the optimiser clear of these regions, but the choice of coordinates matters beyond optimisation.
Laplace's method is accurate to the extent that the log-posterior is close to quadratic, and the parameterisation largely determines how close it is: a Gaussian on the natural scale of a variance puts mass on negative values however well the mode is located, while on the log scale the same posterior is typically much closer to symmetric \autocite{jamil2026approximate}.

\subsubsection{Augmenting the Parameter Table}\label{sec-augmenting}

\texttt{lavaan} stores every model parameter in a \emph{parameter table} (accessible via \texttt{partable()}).
This is a data frame whose rows are individual parameters and whose columns include the left-hand side (\texttt{lhs}), operator (\texttt{op}), right-hand side (\texttt{rhs}), group/level index, whether the parameter is free or fixed, its starting value, and any user-supplied labels or constraints \autocite{rosseel2012lavaan}.
\texttt{INLAvaan} enriches this table, adding four pieces of information to each row:

\begin{enumerate}
\def\labelenumi{\arabic{enumi}.}
\item
  \textbf{Matrix classification.} Each free parameter is tagged with its SEM role (\texttt{lambda}, \texttt{beta}, \texttt{nu}, \texttt{alpha}, \texttt{theta\_var}, \texttt{theta\_cor}, \texttt{psi\_var}, \texttt{psi\_cor}, etc.) by inspecting the operator and whether the variable names relate to observed or latent variables.
\item
  \textbf{Prior assignment.} A prior string (e.g., \texttt{"normal(0,10)"}, \texttt{"gamma(1,.5){[}sd{]}"}, \texttt{"beta(1,1)"}) is attached to each free parameter according to its matrix class, using defaults from \texttt{priors\_for()} that can be overridden globally or per-parameter via a \texttt{prior(...)*\textless{}variable\textgreater{}} syntax in the model string. This follows the same convention as \texttt{blavaan} \autocite{merkle2018blavaan}.
\item
  \textbf{Monotone transforms.} A pair of differentiable bijections \(g_j : \mathcal{D}_j \to \mathbb{R}\) and \(g_j^{-1}: \mathbb{R} \to \mathcal{D}_j\) that map each constrained parameter to the real line and back is assigned based on the matrix type:

  \smallskip
  \begin{center}
  \begin{tabular}{@{}p{0.55\linewidth}p{0.1\linewidth}p{0.1\linewidth}p{0.1\linewidth}@{}}
  \toprule
  Parameter class & $\mathcal{D}_j$ & $g_j$ & $g_j^{-1}$ \\
  \midrule
  Intercepts, loadings, regressions ($\nu_i$, $\Lambda_{ij}$, $\alpha_j$, $B_{jj'}$) & $\mathbb{R}$ & identity & identity \\
  Variances ($\Theta_{ii}$, $\Psi_{jj}$) & $(0,\infty)$ & $\log$ & $\exp$ \\
  Correlations ($\rho$) & $(-1,1)$ & $\operatorname{artanh}$ & $\tanh$ \\
  \bottomrule
  \end{tabular}
  \end{center}
  \smallskip

  Note that for the first two rows, \(g_j^{-1}(\vartheta_j) = x_j\) directly. However, for off-diagonal covariance entries, \(g_j^{-1}\) recovers first the correlation \(\rho_{jk}\), and the natural parameter must be reconstructed as \(x_j = \sigma_j \sigma_k \rho_{jk}\) (see \textbf{Jacobian assembly} below).
  The first and second derivatives of \(g_j^{-1}\) are also stored, enabling exact Jacobian adjustments in the gradient (discussed below) without symbolic differentiation.
\item
  \textbf{Starting values in the unconstrained space.} \texttt{lavaan}'s default starting values---loadings via FABIN3 \autocite{hagglund1982factor}, residual variances at half the observed marginal variance, latent variances at \(0.05\), and all covariances and regression coefficients at zero---must be translated to the unconstrained scale before optimisation.
  \texttt{INLAvaan} adopts these defaults directly: for element-wise parameters, \(\vartheta_j^{\text{init}} = g_j(x_j^{\text{init}})\); for covariance entries, the starting covariance is first decomposed into its standard-deviation and correlation components before applying the respective \(g_j\).
  Users may override starting values via the \texttt{start} argument or \texttt{start(...)} syntax in the model string, always on the natural scale \(\mathbf{x}\), and the translation to \(\boldsymbol\vartheta\) is handled internally.
\end{enumerate}

\begin{refremark}[Positive definiteness]
Each \(\tanh\) map keeps one correlation in \((-1,1)\).
It does not keep a block of several correlations positive definite, such as the residual correlations among \(y_2\), \(y_4\), \(y_6\), and \(y_8\) in the Political Democracy model.
\texttt{INLAvaan} handles this numerically: wherever the implied covariance matrix \(\boldsymbol\Sigma(\mathbf x)\) is not positive definite, the log-likelihood is set to \(-10^{40}\), so the optimiser and the marginal scans stay where the likelihood is defined.
A joint prior and parameterisation for each block would remove the need for this guard.

\label{rem-pd-blocks}

\end{refremark}

Other columns in the \texttt{partable} are left unchanged, so the augmented table retains all of \texttt{lavaan}'s existing information about parameter labels, equality constraints, and group/level structure.
For the Political Democracy model, the first few rows of the augmented table look like this:

\begin{Shaded}
\begin{Highlighting}[]
\NormalTok{R}\SpecialCharTok{\textgreater{}} \FunctionTok{str}\NormalTok{(}\FunctionTok{partable}\NormalTok{(fit), }\AttributeTok{vec.len =} \DecValTok{3}\NormalTok{)}
\end{Highlighting}
\end{Shaded}

\begin{verbatim}
Classes 'lavaan.data.frame' and 'data.frame':   34 obs. of  25 variables:
 $ id         : int  1 2 3 4 5 6 7 8 ...
 $ lhs        : chr  "ind60" "ind60" "ind60" ...
 $ op         : chr  "=~" "=~" "=~" ...
 $ rhs        : chr  "x1" "x2" "x3" ...
 $ user       : int  1 1 1 1 1 1 1 1 ...
 $ block      : int  1 1 1 1 1 1 1 1 ...
 $ group      : int  1 1 1 1 1 1 1 1 ...
 $ free       : int  0 1 2 0 3 4 5 0 ...
 $ ustart     : num  1 NA NA 1 NA NA NA 1 ...
 $ exo        : int  0 0 0 0 0 0 0 0 ...
 $ label      : chr  "" "" "" ...
 $ plabel     : chr  ".p1." ".p2." ".p3." ...
 $ start      : num  1 2.19 1.82 1 ...
 $ est        : num  1 2.22 1.84 1 ...
 $ mat        : chr  "lambda" "lambda" "lambda" ...
 $ prior      : chr  NA "normal(0,10)" "normal(0,10)" ...
 $ g          : chr  "function (x) x" "function (x) x" "function (x) x" ...
 $ g_prime    : chr  "function (x) 1" "function (x) 1" "function (x) 1" ...
 $ ginv       : chr  "function (x) x" "function (x) x" "function (x) x" ...
 $ ginv_prime : chr  "function (x) 1" "function (x) 1" "function (x) 1" ...
 $ ginv_prime2: chr  "function (x) 0" "function (x) 0" "function (x) 0" ...
 $ parstart   : num  1 2.19 1.82 1 ...
 $ names      : chr  "ind60=~x1" "ind60=~x2" "ind60=~x3" ...
 $ se         : num  NA 0.148 0.156 NA ...
 $ par        : num  1 2.21 1.84 1 ...
\end{verbatim}

\subsubsection{Downstream Consumers of the Augmented Table}\label{downstream-consumers-of-the-augmented-table}

The enriched parameter table is the central data structure consulted at every computation.
This subsection summarises the main ways the augmented columns are consumed.

\textbf{Prior evaluation.}
The prior strings from step 2 are parsed once into pairs of R functions---a log-density and its gradient---each accepting the unconstrained parameter value as input.
Both are in closed form for the three supported families (normal, gamma, and beta).
The matrix classification from step 1 determines which family-transform combination applies, and serves as a look-up key whenever the code needs to distinguish, say, a variance parameter from a loading, or more importantly, a covariance parameter from a variance parameter (see next point).

\textbf{Jacobian assembly.}
The monotone transforms from step 3 supply the diagonal entries of the Jacobian \(\partial\mathbf{x}/\partial\boldsymbol\vartheta\), where \(\mathbf{x}\) is the natural parameter vector consumed by \texttt{lavaan}'s likelihood (see next point).
For element-wise transforms, the derivative \((g_j^{-1})'(\vartheta_j)\) is read directly from the stored first-derivative function.
Covariance parameters are the exception: each off-diagonal entry \(\Psi_{jk}\) or \(\Theta_{jk}\) depends jointly on two log-standard-deviations and a Fisher-transformed correlation, introducing off-diagonal Jacobian entries that couple the three unconstrained parameters.
\texttt{INLAvaan} assembles the full (sparse) Jacobian analytically,
with diagonal entries from the stored derivatives and
off-diagonal entries from the product rule applied to the
\(\sigma_j \sigma_k \rho_{jk}\) factorisation.

\textbf{Log-posterior and gradient.}
With the Jacobian \(\mathcal{J}\) and analytic prior gradient in hand, the unconstrained gradient is assembled by the chain rule (see Equation~\ref{eq-chain-rule-grad} in Section~\ref{sec-mode-hessian}), where \texttt{lavaan}'s closed-form ML gradient is multiplied by \(\mathcal{J}^\top\) and the prior gradient is added.
No separate likelihood implementation exists inside \texttt{INLAvaan}: the data-side computation is performed by \texttt{lavaan}, with the adjustment for models without a mean structure described in Section~\ref{sec-preliminaries}.

\textbf{Equality constraints.}
When the model includes equality constraints among parameters (e.g., fixing two loadings to be equal), the augmented parameter table's \texttt{free} and \texttt{label} columns are inspected to identify which parameters should be treated as identical.
\texttt{lavaan}'s \texttt{ceq.simple.K} projection matrix then maps from the reduced parameter space (one unique value per constrained group) to the full space \autocite{rosseel2015lavaan}.
All subsequent computations---optimisation, Hessian, VB correction, marginal profiling---operate in this reduced space, with unpacking applied only when evaluating the log-likelihood.

\textbf{Reporting on the natural scale.}
After all inference is performed in the unconstrained space, the element-wise inverse transforms \(g_j^{-1}\) recover parameters on their constrained scales (loadings, variances, and correlations).
\texttt{pars\_to\_x()} then reconstructs any off-diagonal covariance entries as \(\sigma_j\sigma_k\rho_{jk}\), restoring \texttt{lavaan}'s natural parameterisation \(\mathbf{x}\).

\begin{tcolorbox}[enhanced jigsaw, arc=.35mm, bottomrule=.15mm, bottomtitle=1mm, breakable, colback=white, colbacktitle=quarto-callout-note-color!10!white, colframe=quarto-callout-note-color-frame, coltitle=black, left=2mm, leftrule=.75mm, opacityback=0, opacitybacktitle=0.6, rightrule=.15mm, title=\textcolor{quarto-callout-note-color}{\faInfo}\hspace{0.5em}{Implication for Users}, titlerule=0mm, toprule=.15mm, toptitle=1mm]

Priors, like starting values, are stated on a scale that SEM users can read: normal priors on loadings, regressions, and intercepts (parameterised by their standard deviation), gamma priors on standard deviations (\texttt{{[}sd{]}}, the default) or precisions (\texttt{{[}prec{]}}), and beta priors on correlations, including the correlation behind each free covariance.
The \texttt{priors\_for()} interface changes a whole class in one call (e.g., \texttt{priors\_for(lambda\ =\ "normal(0,0.1)")} tightens all loading priors), while \texttt{prior(...)} in the model string overrides a single parameter.
The \texttt{Prior} column of \texttt{summary()} lists the prior used for every parameter, so the defaults can be checked before interpreting results.

\end{tcolorbox}

\subsection{Posterior Mode and Curvature}\label{sec-mode-hessian}

With the parameter space established, the unconstrained and unnormalised log-posterior is \(\mathcal{L}(\boldsymbol\vartheta) = \log\pi(\mathbf{y} \mid \mathbf{x}(\boldsymbol\vartheta)) + \log\pi(\boldsymbol\vartheta)\), where \(\mathbf{x}(\boldsymbol\vartheta)\) denotes the full mapping from unconstrained parameters to \texttt{lavaan}'s natural vector (element-wise \(g_j^{-1}\) plus covariance reconstruction).
Its gradient is assembled by the chain rule,
\begin{equation}\protect\phantomsection\label{eq-chain-rule-grad}{
\nabla_{\boldsymbol{\vartheta}} \, \mathcal{L}(\boldsymbol{\vartheta})
= \mathcal{J}(\boldsymbol{\vartheta})^\top \,\nabla_{\mathbf{x}} \log\pi(\mathbf{y} \mid \mathbf{x})
\;+\; \nabla_{\boldsymbol{\vartheta}} \log \pi(\boldsymbol{\vartheta}),
}\end{equation}
where \(\mathcal{J}\) is the Jacobian assembled in Section~\ref{sec-param-space}, \(\nabla_{\mathbf{x}} \log\pi(\mathbf{y} \mid \mathbf{x})\) is \texttt{lavaan}'s own closed-form ML gradient, and \(\nabla_{\boldsymbol{\vartheta}} \log \pi(\boldsymbol{\vartheta})\) is the prior gradient.
With this, \texttt{INLAvaan} locates the mode \(\boldsymbol\vartheta^* = \arg\max_{\boldsymbol\vartheta} \mathcal{L}(\boldsymbol\vartheta)\) using R's built-in nonlinear minimisation routine \texttt{nlminb()}.

At the mode, the negative Hessian \(\mathbf{H} = -\nabla^2\mathcal{L}(\boldsymbol\vartheta^*)\) is the precision matrix of the Laplace approximation \(\N_m(\boldsymbol\vartheta^*, \mathbf{H}^{-1})\).
\texttt{INLAvaan} computes \(\mathbf{H}\) via a custom central-difference Jacobian of the analytic gradient (\texttt{fast\_jacobian()}): for each coordinate \(j = 1,\ldots,m\), the gradient is evaluated at \(\boldsymbol\vartheta^* \pm h\mathbf{e}_j\) and differenced, costing \(2m\) gradient evaluations.
This is about a quarter of the \(8m+1\) evaluations of the default Richardson extrapolation in \texttt{numDeriv::jacobian()} \autocite{gilbert2019numderiv}, and is sufficiently accurate for a local curvature estimate used for the Laplace approximation.

\subsubsection{Convergence Diagnostics}\label{convergence-diagnostics}

Monitoring convergence is more consequential here than in frequentist SEM, as the Hessian \(\mathbf{H}\) is not merely a standard-error device but the foundation of the entire subsequent approximation pipeline.
The VB correction, marginal profiling, and copula sampling all rest on \(\mathbf{H}^{-1}\) and cannot proceed without a positive-definite \(\mathbf{H}\), so the fit then stops with an error.
To check convergence, call \texttt{diagnostics()}:

\begin{Shaded}
\begin{Highlighting}[]
\NormalTok{R}\SpecialCharTok{\textgreater{}} \FunctionTok{diagnostics}\NormalTok{(fit)}
\end{Highlighting}
\end{Shaded}

\begin{verbatim}
             npar             nsamp         converged        iterations 
               31              1000                 1                65 
         grad_inf      grad_inf_rel           grad_l2    mode_shift_max 
         6.58e-04          1.31e-03          8.89e-04          1.29e-04 
        hess_cond        vb_applied      vb_shift_max     vb_kld_global 
         2.39e+02                 1            0.5221            0.5967 
          kld_max          kld_mean       vb_mcse_max      vb_mcse_mean 
           0.1363            0.0286            0.0853            0.0314 
         nmad_max         nmad_mean scan_end_mass_max 
           0.0434            0.0076          3.62e-03 
\end{verbatim}

We now go through how this output establishes whether the returned point is a valid mode.
\texttt{converged} (binary) and \texttt{iterations} come from \texttt{nlminb()}.
\texttt{grad\_inf}, \texttt{grad\_inf\_rel}, and \texttt{grad\_l2} are the \(\ell^\infty\) norm, the \(\ell^\infty\) norm relative to the size of each parameter, and the \(\ell^2\) norm of the gradient of the log-posterior at that point, which should be close to zero.
Their size depends on the units of the parameters, however, so they have no universal threshold.
\texttt{mode\_shift\_max} removes this dependence: it is the largest element of the Newton step \(\mathbf{H}^{-1}\nabla\mathcal{L}(\boldsymbol\vartheta^*)\) in posterior standard deviations, i.e., how far one more Newton step would move a parameter relative to its posterior uncertainty, and \texttt{INLAvaan} warns when it exceeds \(0.1\).
Positive definiteness of \(\mathbf{H}\) is checked by its Cholesky factorisation, and \texttt{hess\_cond}, the ratio of its largest to its smallest eigenvalue, guards against near-singularity, with a warning above \(10^8\).
The remaining fields summarise the quality of the approximation and are discussed in the sections that follow.

\begin{tcolorbox}[enhanced jigsaw, arc=.35mm, bottomrule=.15mm, bottomtitle=1mm, breakable, colback=white, colbacktitle=quarto-callout-note-color!10!white, colframe=quarto-callout-note-color-frame, coltitle=black, left=2mm, leftrule=.75mm, opacityback=0, opacitybacktitle=0.6, rightrule=.15mm, title=\textcolor{quarto-callout-note-color}{\faInfo}\hspace{0.5em}{Implication for Users}, titlerule=0mm, toprule=.15mm, toptitle=1mm]

To help users spot an invalid mode, \texttt{INLAvaan} warns at the end of the fit when (a) the optimiser did not converge; (b) the mode shift exceeds 0.1 posterior standard deviations; or (c) the condition number of the Hessian exceeds \(10^8\).
These thresholds are package defaults, not calibrated cut-offs, so the full \texttt{diagnostics()} output lets users judge the mode for themselves, and \texttt{diagnostics(fit,\ type\ =\ "param")} breaks it down by parameter.
Warnings (a) and (b) can often be resolved through the \texttt{control} argument, which is passed to \texttt{nlminb()} (see \texttt{?nlminb} for the available settings).
For instance, \texttt{control\ =\ list(iter.max\ =\ 5000,\ eval.max\ =\ 5000)} raises the iteration and evaluation limits, and \texttt{control\ =\ list(rel.tol\ =\ 1e-12)} tightens the convergence tolerance.
A warning (c), or a mode shift that persists under a tighter tolerance, usually points to a weakly identified parameter.

\end{tcolorbox}

\subsection{Variational Bayes Location Correction}\label{sec-vb}

\begin{figure}[H]

\centering{

\includegraphics[width=1\linewidth,height=\textheight,keepaspectratio]{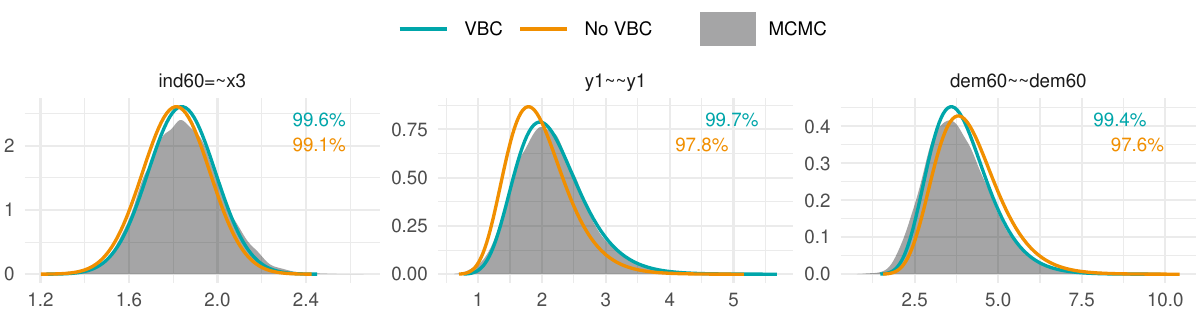}

}

\caption{\label{fig-vbc}A clear improvement in approximation quality of the Gaussian posterior marginals (\texttt{marginal\_method\ =\ "marggaus"}) is seen due to the VB mean-shift correction (VBC) on three selected parameters from the Political Democracy example: a factor loading (\texttt{ind60=\textasciitilde{}x3}), a residual variance (\texttt{y1\textasciitilde{}\textasciitilde{}y1}), and a latent variance (\texttt{dem60\textasciitilde{}\textasciitilde{}dem60}). Percentages are Jensen--Shannon similarities to MCMC (higher is better).}

\end{figure}

The Laplace approximation centres the Gaussian at the posterior mode \(\boldsymbol\vartheta^*\) in the unconstrained space.
The centring is adequate when the log-posterior is approximately quadratic, a condition that typically holds for loadings and regressions at moderate-to-large sample sizes, but is not guaranteed for any parameter class.
Such departures are most pronounced for variance components (see Figure~\ref{fig-vbc}), yet can affect any marginal when the sample is small, the model is weakly identified, or prior-data conflict is present.
In all such cases the posterior is skewed and the mode can be displaced from the mean in either direction, propagating bias to marginal summaries \autocite{jamil2026approximate}.

\protect\phantomsection\label{cell-fig-vbc}

\vspace{-0.5em}

\texttt{INLAvaan} corrects for this by estimating a VB mean shift \autocite[Eq. 12]{jamil2026approximate}.
Holding the variance-covariance matrix \(\boldsymbol\Omega=\mathbf H^{-1}\) of the Gaussian approximation fixed, the shift \(\hat{\boldsymbol\delta}\) is found by maximising a quasi-Monte Carlo (QMC) approximation to the expected unnormalised log-posterior,
\begin{equation}\protect\phantomsection\label{eq-vb-elbo}{
\hat{\boldsymbol\delta}
= \argmax_{\boldsymbol\delta} \left\{ \frac{1}{B+1}\sum_{b=0}^{B} \mathcal{L} \left(\boldsymbol\vartheta^* + \boldsymbol\delta + \mathbf{L}\mathbf{u}_b\right) \right\}
\approx \argmax_{\boldsymbol\delta} \int \mathcal{L}(\boldsymbol\vartheta) \, \phi(\boldsymbol\vartheta; \boldsymbol\vartheta^* + \boldsymbol\delta, \boldsymbol\Omega) \, d\boldsymbol\vartheta
}\end{equation}
where \(\mathbf{u}_0 = \mathbf{0}\) contributes the mode itself, \(\mathbf{u}_1,\ldots,\mathbf{u}_B\) are scrambled Owen--Sobol points \autocite{owen1998scrambling,joe2008constructing} mapped to \(\N(\mathbf{0}, \mathbf{I})\), \(B = 64\) by default (argument \texttt{n\_qmc}), and \(\mathbf{L}\) is the lower Cholesky factor of \(\boldsymbol\Omega\).
Each term \(\boldsymbol\delta + \mathbf{L}\mathbf{u}_b\) is therefore approximately a draw from \(\N( \boldsymbol\delta, \boldsymbol\Omega)\), giving each evaluation point the appropriate posterior spread around \(\boldsymbol\vartheta^*\).

The package ships a pre-computed Sobol table generated from \texttt{spacefillr}'s \autocite{morgan2025spacefillr} \texttt{generate\_sobol\_owen\_set()} function, falling back to the \texttt{qrng} package \autocite{hofert2026qrng} for high-dimensional models.
Crucially, the gradient of (\ref{eq-vb-elbo}) with respect to \(\boldsymbol\delta\) is a sample average of \(\nabla_{\boldsymbol\vartheta}\mathcal{L}\), so the same analytic gradient from Section~\ref{sec-mode-hessian} is reused directly at each quadrature point with no additional implementation.
Since the Newton step for the curvature \(\mathbf{H}\) is a single multiplication by \(\boldsymbol\Omega\), \texttt{INLAvaan} solves for \(\hat{\boldsymbol\delta}\) by the fixed-point iteration \(\boldsymbol\delta \leftarrow \boldsymbol\delta + \boldsymbol\Omega\,\overline{\nabla\mathcal{L}}\), where \(\overline{\nabla\mathcal{L}}\) is the average gradient over the nodes; in the examples of this paper it converges in 2 to 14 sweeps.
If a step is too large, the gradient is not finite, or the iteration does not converge, the solver falls back to \texttt{nlminb()} on the whitened coordinates \(\mathbf{d} = \mathbf{L}^{-1}\boldsymbol\delta\), which rescale every search direction to unit variance.

The shift \(\hat{\boldsymbol\delta}\) bringing the original Laplace Gaussian \(q_0 \equiv \N(\boldsymbol\vartheta^*, \boldsymbol\Omega)\) to \(q_{\hat\delta}\equiv \N(\boldsymbol\vartheta^* + \hat{\boldsymbol\delta}, \boldsymbol\Omega)\) varies considerably across parameters, from negligible for well-identified loadings to several tenths of a posterior standard deviation for variance components.
\texttt{diagnostics()} records whether the correction was applied (\texttt{vb\_applied}) and three diagnostics of it, given per parameter by \texttt{diagnostics(fit,\ type\ =\ "param")} and summarised across parameters in the default output:

\begin{itemize}
\item
  \textbf{Shift:} The \(j\)-th Gaussian marginals of \(q_0\) and \(q_{\hat\delta}\) are both \(\N(\cdot, \Omega_{jj})\), differing only in location by \(\hat\delta_j\).
  The shift in posterior standard deviations, \(d_j = \hat\delta_j/\Omega_{jj}^{1/2}\) (\texttt{vb\_shift\_sigma}, with largest absolute value \texttt{vb\_shift\_max}), reads like a \(z\)-score: the average absolute shift of 0.21 in the worked example moves the centre from the 50th to the 58th percentile of the Laplace marginal of \(q_0\).
  \texttt{INLAvaan} warns when \texttt{vb\_shift\_max} exceeds one standard deviation.
  Across 1,046 simulated fits in \textcite[Supplementary Material, Table S1]{jamil2026approximate}, the VB shift was the most strongly associated of the three approximation diagnostics with the worst-case error of the posterior means against MCMC.
\item
  \textbf{KL divergence:} For two univariate Gaussians with equal variances, the KL divergence reduces to \((\Delta\mu)^2/(2\sigma^2)\), so the per-parameter \(\text{KLD}_j = d_j^2/2\) (\texttt{kld}, summarised by \texttt{kld\_max} and \texttt{kld\_mean}) is the shift on another scale (an information scale).
  The joint Gaussians \(q_0\) and \(q_{\hat\delta}\) likewise differ only in location, so their KL divergence is
  \[
  \text{KLD}_\text{global} = D_\mathrm{KL}(q_0 \| q_{\hat\delta}) = \tfrac12\,\hat{\boldsymbol\delta}^\top \boldsymbol\Omega^{-1} \hat{\boldsymbol\delta}
  \qquad (\texttt{vb\_kld\_global}),
  \]
  the joint counterpart of \(\text{KLD}_j\).
  To leading order it also equals the improvement in approximation quality achieved by the VB shift.
  Since \(D_\mathrm{KL}(q \| \pi) = \mathrm{const} - \operatorname{E}_q[\mathcal{L}(\boldsymbol\vartheta)]\) with the constant independent of \(\boldsymbol\delta\), this improvement is
  \[
  \Delta D_\mathrm{KL}
  = D_\mathrm{KL}(q_0 \| \pi) - D_\mathrm{KL}(q_{\hat\delta} \| \pi)
  = \operatorname{E}_{q_{\hat\delta}}\!\big[\mathcal{L}(\boldsymbol\vartheta)\big] - \operatorname{E}_{q_0}\!\big[\mathcal{L}(\boldsymbol\vartheta)\big]
  \geq 0.
  \]
  The VB shift adds \(\text{KLD}_\text{global}\) to the Laplace log-evidence in (\ref{eq-marglik}) as its location term \autocite[Sec. 3.1]{jamil2026approximate}.
\item
  \textbf{Quadrature error:} The shift solves an integral approximated over the \(B\) QMC nodes, so it carries an integration error of its own.
  \texttt{INLAvaan} estimates this error by splitting the nodes into two halves, solving for the shift on each, and taking half the difference between the two solutions, expressed in posterior standard deviations (\texttt{vb\_mcse\_sigma}, summarised by \texttt{vb\_mcse\_max} and \texttt{vb\_mcse\_mean}).
  It reads as an ``error bar'' on \(d_j\), in that a value of 0.05 means the corrected centre could move by about 0.05 posterior standard deviations through the choice of nodes alone.
\end{itemize}

\begin{tcolorbox}[enhanced jigsaw, arc=.35mm, bottomrule=.15mm, bottomtitle=1mm, breakable, colback=white, colbacktitle=quarto-callout-note-color!10!white, colframe=quarto-callout-note-color-frame, coltitle=black, left=2mm, leftrule=.75mm, opacityback=0, opacitybacktitle=0.6, rightrule=.15mm, title=\textcolor{quarto-callout-note-color}{\faInfo}\hspace{0.5em}{Implication for Users}, titlerule=0mm, toprule=.15mm, toptitle=1mm]

The VB correction is on by default (\texttt{vb\_correction\ =\ TRUE}).
It takes a modest share of the fitting time and often improves the posterior means substantially \autocite[Sec. 4.3]{jamil2026approximate}, so there is rarely a reason to disable it.
In essence, the VB shift indicates how close each posterior is to the symmetric, Gaussian shape it takes when the data are highly informative about that parameter.
A shift above one standard deviation triggers a warning and points to weak identification or prior-data conflict, so check the constraints and prior of that parameter, or confirm the result with MCMC.
The quadrature error qualifies the shift, and a \texttt{vb\_mcse\_max} that is not small next to the shifts means the node set is too coarse, so raise \texttt{n\_qmc} (fitting takes longer).
\texttt{summary(fit,\ vb\_shift\ =\ TRUE)} and \texttt{summary(fit,\ kld\ =\ TRUE)} print the shift and the KLD as columns.

\end{tcolorbox}

\subsection{Marginal Posterior Approximation}\label{sec-marginals}

The VB-corrected Gaussian \(q_{\hat\delta}\) captures the location and scale of the posterior, but not its asymmetry, so each marginal \(\pi(\vartheta_j \mid \mathbf{y})\) must be approximated directly.
The standard Laplace approximation to each marginal \autocite{tierney1986accurate} integrates out the remaining \(m-1\) parameters by a further Laplace step, which entails re-evaluating the \((m-1)\times(m-1)\) conditional Hessian determinant at each candidate value of \(\vartheta_j\), a cost that grows rapidly with \(m\).
\texttt{INLAvaan} instead profiles the marginals \(\{\pi(\vartheta_j \mid \mathbf{y})\}_{j=1}^m\) via the three-step procedure of \textcite{jamil2026approximate}, which consists of axis scanning along the conditional mean direction, followed by a gradient-based
volume correction, and skew-normal fitting (described in the next subsections).

\begin{figure}[H]

\centering{

\includegraphics[width=1\linewidth,height=\textheight,keepaspectratio]{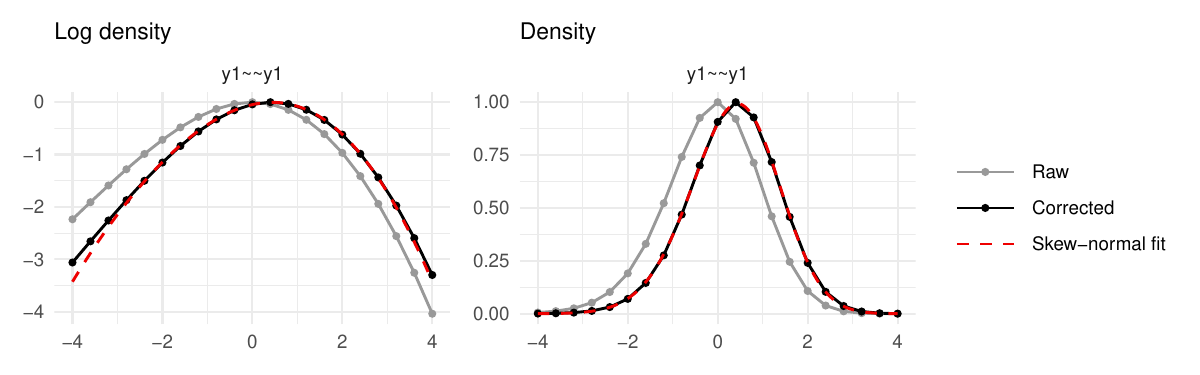}

}

\caption{\label{fig-margskewnorm}Log-profile (left) and corresponding density (right) for the residual variance \texttt{y1\textasciitilde{}\textasciitilde{}y1} in the Political Democracy model. The uncorrected raw log-profile peaks at the mode; the volume correction tilts it, moving the marginal about 0.4 posterior standard deviations to the right. The skew-normal fit (red dashed) closely tracks the corrected log-profile, but may deviate in the tails where the profile is less informative.}

\end{figure}

Beginning with the axis scan for parameter \(j\), the log-posterior is evaluated at a grid of 21 standardised points \(z_k \in [-4, 4]\) along the \emph{scan direction} \(\mathbf{v}_j = \boldsymbol\Omega_{\cdot,j} / \sqrt{\Omega_{jj}}\); i.e., at \(\boldsymbol\vartheta^* + \mathbf{v}_j z_k\) for \(k = 1,\ldots,21\) \autocite[Eq. 13, Sec. 3.2.1]{jamil2026approximate}.
In our testing (see Appendix A), accuracy stops improving beyond about nine points, and 21 points add about 9\% to the fitting time, so the default keeps a margin of safety; Figure~\ref{fig-margskewnorm} shows such a grid for one parameter.
Each evaluation requires one call to the joint log-posterior, which is dominated by the \(O(p^3)\) cost of the model-implied covariance \(\boldsymbol\Sigma\).
The resulting raw log-profile is a 21-point record of the log-posterior \emph{height} along the scan grid, approximating \(\log\pi(\vartheta_j \mid \mathbf{y})\) up to an additive constant.

\protect\phantomsection\label{cell-fig-margskewnorm}

\subsubsection{Volume Correction}\label{sec-volume-correction}

As detailed in \textcite{jamil2026approximate}, the raw log-profile mixes the marginal \(\pi(\vartheta_j \mid \mathbf y)\) with the height of the conditional density of the remaining parameters, which changes along the scan as their spread changes, and so misplaces the marginal's location.
The correction takes the form of a linear tilt: the adjusted log-profile is \(\mathcal{L}_j^{\text{adj}}(z_k) = \mathcal{L}_j^{\text{raw}}(z_k) + \gamma_{j}' \, z_k\), where \(\gamma_{j}'\) captures the rate of change of the log-determinant of the conditional Hessian along the scan direction.

\texttt{INLAvaan} provides four methods for computing \(\gamma_{j}'\), selectable via the \texttt{marginal\_correction} argument:

\vspace{-0.15em}

{\def\LTcaptype{none} % do not increment counter
\begin{longtable}[]{@{}
  >{\raggedright\arraybackslash}p{(\linewidth - 4\tabcolsep) * \real{0.2143}}
  >{\raggedright\arraybackslash}p{(\linewidth - 4\tabcolsep) * \real{0.2143}}
  >{\raggedright\arraybackslash}p{(\linewidth - 4\tabcolsep) * \real{0.5714}}@{}}
\toprule\noalign{}
\begin{minipage}[b]{\linewidth}\raggedright
Method
\end{minipage} & \begin{minipage}[b]{\linewidth}\raggedright
Cost per parameter
\end{minipage} & \begin{minipage}[b]{\linewidth}\raggedright
Computation
\end{minipage} \\
\midrule\noalign{}
\endhead
\bottomrule\noalign{}
\endlastfoot
\texttt{"shortcut"} (default) & \(2m + 2\) gradient evals & Central-difference trace of \(\partial\mathbf{H}/\partial z_j\), plus a Schur complement correction \\
\texttt{"shortcut\_fd"} & \(m + 2\) gradient evals & Forward-difference variant of the above, at about half the cost but less accurate in principle \\
\texttt{"hessian"} & \(4m\) gradient evals & Full Hessian at two shifted points via central finite difference \\
\texttt{"none"} & \(0\) & No correction: the raw log-profile is fitted as is (for benchmarking only) \\
\end{longtable}
}

\vspace{-0.85em}

For most applied models, the default \texttt{"shortcut"} strikes the right balance.
In the ablation study of \textcite[Table D1]{jamil2026approximate}, it was as accurate as \texttt{"hessian"}.
\texttt{"shortcut\_fd"} roughly halves the cost of the default but is less accurate in principle, since it uses forward differences throughout, so we reserve it for large models where fitting time matters, such as the 256-parameter model of Section~\ref{sec-circfa}.
\texttt{"hessian"} then serves as a reference check, and \texttt{"none"} only as a benchmark.
To check whether the marginals depend on the numerical volume-correction method, refit with \texttt{"hessian"} and compare.

\subsubsection{Fitting the Skew-Normal}\label{fitting-the-skew-normal}

The corrected 21-point log-profile is fitted to a skew-normal density \(\text{SN}(\xi_j, \omega_j, \alpha_j)\) by minimising a weighted sum of squared residuals between the log-profile and the log SN density, with weights proportional to the profile height and zero weight where the log-profile lies more than 6 below its peak.
This naturally concentrates the fit on the central region of the posterior where the density mass lies.
An additional free normalisation constant \(c_j\) is included as a fourth parameter alongside \((\xi_j, \omega_j, \alpha_j)\), since the log-profile is only defined up to an additive constant.
This allows the SN to fit the shape of the profile without requiring it to integrate exactly to one.

Starting values are obtained by integrating the exponentiated profile with the trapezoidal rule to extract empirical moments, solving for the \((\xi_j, \omega_j, \alpha_j)\) triple via the skewness-delta relation \autocite[Sec. 2.2]{jamil2026approximate}, which avoids flat-start convergence failures.
The objective function is supplied to \texttt{nlminb()} together with its analytic gradient and Hessian, ensuring fast and reliable convergence for every profile in the loop.

The fitted parameters \((\hat\xi_j, \hat\omega_j, \hat\alpha_j)\) are used in two downstream steps: summary statistics (mean and SD by Gauss--Hermite quadrature on the natural scale, quantiles from \texttt{qsnorm\_fast()}), and copula sampling via the fast inverse CDF \texttt{qsnorm\_fast()} (discussed below in Section~\ref{sec-qsnormfast}).
The shape parameter \(\hat\alpha_j\) (\texttt{alpha} in \texttt{diagnostics(fit,\ type\ =\ "param")}, on the unconstrained scale) is a useful indicator of the direction and magnitude of skewness in each marginal, with \(\hat\alpha_j = 0\) corresponding to an exact Gaussian.

\subsubsection{Parallelisation}\label{parallelisation}

The per-parameter profiling and fitting loop is embarrassingly parallel: each parameter's computation depends only on the shared mode, covariance, and scan directions.
When \texttt{cores\ \textgreater{}\ 1} (or \texttt{cores\ =\ NULL} with \(m > 120\)), \texttt{INLAvaan} distributes the loop across cores, forking with \texttt{parallel::mclapply()} where that is safe and otherwise using a cluster of separate R processes inside integrated development environments such as RStudio and Positron (on Windows, the marginals currently run serially).
For the 256-parameter circumplex model in Section~\ref{sec-circfa}, this reduces marginal fitting from about 47 seconds (serial) to 8 seconds on 14 cores.

\subsubsection{Fit Quality Diagnostics}\label{fit-quality-diagnostics}

Because the fitting objective emphasises the central mass of the profile, the SN may fit the peak well while diverging in the tails---a discrepancy that would not be visible from a central-tendency summary alone.
The normalised maximum absolute deviation (NMAD) is designed to catch this:
\[
\text{NMAD}_j
= \frac{\max_k\bigl|\exp(\mathcal{L}_j^\text{adj}(z_k)) - \hat{f}_j(z_k)\bigr|}{\max_k \exp(\mathcal{L}_j^\text{adj}(z_k))},
\]
where \(\hat{f}_j \equiv f_{\text{SN}}(\cdot\,;\hat\xi_j,\hat\omega_j, \hat\alpha_j)\) is the fitted SN density evaluated at the same grid points.
Being normalised by the profile peak, NMAD is a dimensionless, per-parameter index of relative goodness-of-fit.
A value of \(0\) is a perfect match, and values below \(0.05\) indicate that the fitted SN is a trustworthy representation of the marginal profile.

Large NMAD values (above \(0.10\), say) almost always signal a tail discrepancy, since the skew-normal family cannot accommodate the curvature present in the tails of the profile (e.g., a bimodal or heavy-shouldered marginal).
In the simulations of \textcite[Sec. 4.4]{jamil2026approximate}, NMAD rose from 0.02--0.05 in benign conditions to 0.10 under weak measurement.
As the SN fit is weighted towards the centre, the central estimates (mean, SD) may still be acceptable, but credible intervals at the 2.5\% or 97.5\% level should be treated with caution.
Per-parameter NMAD values are reported in the output of \texttt{summary()} or \texttt{diagnostics()}.
For reference, the NMAD value for \texttt{y1\textasciitilde{}\textasciitilde{}y1} in the Political Democracy example (Figure~\ref{fig-margskewnorm}) is 0.009, indicating minimal discrepancy.

NMAD checks the fit only at the scan points, which span four Laplace standard deviations either side of the mode, so it cannot see how much of the fitted marginal lies beyond them.
The scan-endpoint mass (\texttt{scan\_end\_mass}),
\[
\text{EM}_j = \hat F_j\big(\vartheta_j^* - 4\,\Omega_{jj}^{1/2}\big) + 1 - \hat F_j\big(\vartheta_j^* + 4\,\Omega_{jj}^{1/2}\big),
\]
where \(\hat F_j\) is the fitted SN distribution function, measures this mass.
Where it is large, the tail quantiles, and hence the credible limits, rest on extrapolation.
For reference, a Gaussian marginal gives \(2\Phi(-4) \approx 6 \times 10^{-5}\), and no parameter of the Political Democracy model exceeds 0.004 (\texttt{scan\_end\_mass\_max}).

\begin{tcolorbox}[enhanced jigsaw, arc=.35mm, bottomrule=.15mm, bottomtitle=1mm, breakable, colback=white, colbacktitle=quarto-callout-note-color!10!white, colframe=quarto-callout-note-color-frame, coltitle=black, left=2mm, leftrule=.75mm, opacityback=0, opacitybacktitle=0.6, rightrule=.15mm, title=\textcolor{quarto-callout-note-color}{\faInfo}\hspace{0.5em}{Implication for Users}, titlerule=0mm, toprule=.15mm, toptitle=1mm]

\texttt{INLAvaan} warns when the NMAD of a parameter exceeds 0.1, or when more than 0.05 of its fitted mass lies beyond the scan.
The posterior mean and standard deviation of a flagged parameter may still be accurate, but its 2.5\% and 97.5\% limits should be used with care, and checked with MCMC if the conclusions rest on them.
To see why a parameter was flagged, \texttt{plot(fit,\ type\ =\ "sn\_fit")} (or \texttt{type\ =\ "sn\_fit\_log"} on the log scale) draws the raw and volume-corrected scans together with the fitted skew-normal, as in Figure~\ref{fig-margskewnorm}.

\end{tcolorbox}

\subsection{Joint Posterior Sampling via Gaussian Copula}\label{sec-copula}

The previous stage produces \(m\) fitted marginal CDFs \(\hat F_j = F_{\text{SN}}(\cdot\,;\hat\xi_j,\hat\omega_j, \hat\alpha_j)\) and the Laplace correlation matrix \(\mathbf{R} = \operatorname{cor}(\boldsymbol\Omega)\).
To generate joint posterior samples that respect both the marginal shapes and the posterior dependence structure, \texttt{INLAvaan} uses a Gaussian copula:
\begin{equation}\protect\phantomsection\label{eq-copula}{
\mathbf{z} \sim \N_m(\mathbf{0}, \mathbf{R}), \quad
u_j = \Phi(z_j), \quad
\vartheta_j = \hat F_j^{-1}(u_j), \quad j = 1,\ldots,m.
}\end{equation}
This sampling is performed internally during fitting to produce the posterior summaries reported by \texttt{summary()}.
It is also exposed directly to users via \texttt{sampling()} for custom posterior computations.

\subsubsection{The Fast Skew-Normal Quantile Function}\label{sec-qsnormfast}

The copula requires evaluating \(\hat F_j^{-1}\) for every sample and every parameter---potentially millions of calls.
Standard iterative root-finding of the SN CDF \autocite[e.g., using \texttt{sn::qsn()},][]{azzalini2023package} would be highly inefficient.
\texttt{INLAvaan} instead uses \texttt{qsnorm\_fast()}, an approximation algorithm developed by \textcite{luu2016fast} and adapted by \textcite{rue2021gmrflib} for the GMRFLib C library behind \texttt{R-INLA}.
The method uses a domain decomposition strategy with pre-tabulated polynomial coefficients, achieving \(< 10^{-7}\) relative error without iteration.
This makes the copula sampler fast enough for routine use.

\begin{Shaded}
\begin{Highlighting}[]
\NormalTok{R}\SpecialCharTok{\textgreater{}} \FunctionTok{library}\NormalTok{(microbenchmark)}
\NormalTok{R}\SpecialCharTok{\textgreater{}} \FunctionTok{microbenchmark}\NormalTok{(}
\SpecialCharTok{+}    \AttributeTok{sn =}\NormalTok{ sn}\SpecialCharTok{::}\FunctionTok{qsn}\NormalTok{(}\FloatTok{0.95}\NormalTok{, }\AttributeTok{xi =} \DecValTok{0}\NormalTok{, }\AttributeTok{omega =} \DecValTok{1}\NormalTok{, }\AttributeTok{alpha =} \DecValTok{5}\NormalTok{),}
\SpecialCharTok{+}    \AttributeTok{INLAvaan =}\NormalTok{ INLAvaan}\SpecialCharTok{::}\FunctionTok{qsnorm\_fast}\NormalTok{(}\FloatTok{0.95}\NormalTok{, }\AttributeTok{xi =} \DecValTok{0}\NormalTok{, }\AttributeTok{omega =} \DecValTok{1}\NormalTok{, }\AttributeTok{alpha =} \DecValTok{5}\NormalTok{),}
\SpecialCharTok{+}    \AttributeTok{times =} \DecValTok{100}\NormalTok{L,}
\SpecialCharTok{+}    \AttributeTok{check =} \StringTok{"equal"}  \CommentTok{\# verify outputs are identical}
\SpecialCharTok{+}\NormalTok{  )}
\end{Highlighting}
\end{Shaded}

\begin{verbatim}
Unit: microseconds
     expr     min       lq      mean   median      uq     max neval
       sn 141.409 143.4180 148.20639 144.3405 147.313 261.375   100
 INLAvaan   5.084   5.3505   6.11064   5.6170   6.396  25.748   100
\end{verbatim}

\subsubsection{Sampling-Based Derived Quantities}\label{sampling-based-derived-quantities}

The copula samples are the gateway to all quantities that are nonlinear functions of the parameter vector.
After drawing the \(\texttt{nsamp} \times m\) matrix in the unconstrained \(\boldsymbol\vartheta\) space, each row is mapped to \texttt{lavaan}'s natural scale \(\mathbf{x}\) via \texttt{pars\_to\_x()}.
From this \texttt{x\_samp} matrix, \texttt{INLAvaan} computes:

\begin{itemize}
\item
  \textbf{Covariance parameters}: For parameters that are products of other parameters (i.e., \(\Psi_{jk} = \rho_{jk} \sigma_j \sigma_k\)), the marginal summaries are obtained from the empirical distribution of the product across samples. When \texttt{sn\_fit\_sample\ =\ TRUE}, a skew-normal is fitted to these samples, producing smooth density curves and fast quantiles.
\item
  \textbf{Defined parameters}: Expressions specified via `\texttt{:=}' in the model string (e.g., \texttt{ind\ :=\ a\ *\ b} for indirect effects) are evaluated row-wise on the sample matrix. Again, when \texttt{sn\_fit\_sample\ =\ TRUE}, a parametric SN is fitted, which is especially valuable for products of coefficients whose sampling distribution is typically skewed.
\item
  \textbf{Posterior covariance}: \texttt{vcov()} returns the posterior covariance on the natural scale \(\mathbf{x}\), computed empirically as the sample covariance of the copula draws mapped back through \texttt{pars\_to\_x()}, matching the parameter scale that \texttt{lavaan} users expect.
  As an aside, \texttt{vcov(...,\ type\ =\ "theta")} instead returns the Laplace covariance \(\boldsymbol\Omega\) in the unconstrained \(\boldsymbol\vartheta\) space.
\item
  \textbf{Prior predictive samples}: Setting \texttt{sampling(fit,\ prior\ =\ TRUE)} bypasses the posterior and instead draws each \(\vartheta_j\) from its prior, then propagates these through the generative model (\(\boldsymbol\vartheta \to \boldsymbol\eta \to \mathbf{y}^*\)), enabling prior predictive checks without any additional code.
\end{itemize}

\begin{tcolorbox}[enhanced jigsaw, arc=.35mm, bottomrule=.15mm, bottomtitle=1mm, breakable, colback=white, colbacktitle=quarto-callout-note-color!10!white, colframe=quarto-callout-note-color-frame, coltitle=black, left=2mm, leftrule=.75mm, opacityback=0, opacitybacktitle=0.6, rightrule=.15mm, title=\textcolor{quarto-callout-note-color}{\faInfo}\hspace{0.5em}{Implication for Users}, titlerule=0mm, toprule=.15mm, toptitle=1mm]

A Gaussian copula allows \texttt{INLAvaan} to generate joint posterior samples that respect both the fitted marginal shapes and the posterior dependence structure, without the need for costly MCMC sampling.
This is used internally during fitting to produce posterior summaries, and is also exposed to users via \texttt{sampling(...,\ type\ =\ "lavaan")} for custom posterior computations.
Posterior parameter samples are suitable for computing any posterior functional---standardised coefficients, \(R^2\) decompositions, fit indices---via simple \texttt{apply()}.
Other \texttt{type} options include \texttt{"latent"} and \texttt{"observed"}.
Prior predictive sampling (\texttt{prior\ =\ TRUE}) is also available \autocite{talts2020validating}.

\end{tcolorbox}

\subsection{Post-Hoc Inference}\label{sec-posthoc}

The fitted \texttt{INLAvaan} object, together with the copula sampler, supports a suite of post-hoc quantities, all obtainable from a single model fit.
We organise these into three capabilities.

\subsubsection{Factor Scores and Missing Data Imputation}\label{sec-predict}

For a given posterior draw \(\mathbf x^{(b)}\), the conditional distribution of the latent variables is available in closed form:
\begin{equation}\protect\phantomsection\label{eq-factor-scores}{
\boldsymbol\eta_s \mid \mathbf{y}_s, \mathbf x^{(b)} \sim \N_q \big(\mathbf{m}_s(\mathbf x^{(b)}), \mathbf{V}(\mathbf x^{(b)})\big),
}\end{equation}
where \(\mathbf{m}_s\) and \(\mathbf{V}\) are the closed-form conditional posterior mean and covariance given in \textcite[see their Eq. 8]{jamil2026approximate}.
The \texttt{predict()} method propagates uncertainty by evaluating (\ref{eq-factor-scores}) at each of many posterior draws of \(\mathbf x\):

\begin{itemize}
\tightlist
\item
  \texttt{predict(fit,\ type\ =\ "lv")} returns a list of \texttt{nsamp} matrices, each an \(n \times q\) matrix drawn from (\ref{eq-factor-scores}), giving full posterior distributions over individual-level factor scores for individuals \(s = 1,\ldots,n\).
  \texttt{predict(fit,\ summary\ =\ TRUE)} instead returns the posterior mean, standard deviation, quantiles, and mode of each factor score.
\item
  \texttt{predict(fit,\ type\ =\ "ov")} returns the model-implied conditional means \(\operatorname{E}(\mathbf{y}_s \mid \boldsymbol\eta_s^{(b)}, \mathbf x^{(b)}) = \boldsymbol\nu^{(b)} + \boldsymbol\Lambda^{(b)}\boldsymbol\eta_s^{(b)}\) without residual noise, useful for examining the structural signal stripped of measurement error.
\item
  \texttt{predict(fit,\ type\ =\ "ypred")} adds a residual draw \(\boldsymbol\epsilon_s^{(b)} \sim \N(\mathbf{0}, \boldsymbol\Theta^{(b)})\), producing realisations from the posterior predictive distribution of observed scores.
\item
  \texttt{predict(fit,\ type\ =\ "ymis")} imputes missing observations by sampling from \(Y_{si}^{\text{mis}} \mid \mathbf{y}_{si}^{\text{obs}},\, \mathbf x^{(b)} \sim \N \bigl(\mu_{\text{mis}|\text{obs}}^{(b)}, \sigma^{2\,(b)}_{\text{mis}|\text{obs}}\bigr)\), using, for single-level models, a pattern-based Cholesky scheme that groups observations by their missing-data pattern to avoid redundant matrix inversions. The conditional moments follow the usual partitioned-normal formulae applied to \(\boldsymbol{\mu}(\mathbf x^{(b)})\) and \(\boldsymbol{\Sigma}(\mathbf x^{(b)})\) \autocite[Ch. 5]{enders2022applied}.
\end{itemize}

For multilevel models, each posterior draw \(\mathbf x^{(b)}\) is followed by one exact Gaussian draw of each cluster's between-level values given all of the cluster's rows, and then by draws of the latent variables and the missing values within each level given those values.
This yields posterior distributions over cluster-level latent constructs, with uncertainty reflecting both the within-cluster sampling noise and the between-level structural model.

\subsubsection{Bayesian Fit Indices and Model Comparison}\label{sec-fitindices}

\texttt{INLAvaan} provides three complementary model-assessment mechanisms, described below.

\textbf{Bayes factors.} The Laplace approximation to the log model evidence is stored in every fitted object:
\begin{equation}\protect\phantomsection\label{eq-marglik}{
\log \tilde\pi(\mathbf{y}) =\mathcal{L}(\boldsymbol\vartheta^*) + \tfrac{m}{2}\log 2\pi + \sum_{k=1}^m \log L_{kk} + \text{KLD}_{\text{global}},
}\end{equation}
where \(L_{kk}\) are the diagonal entries of the Cholesky factor of \(\boldsymbol\Omega\) and \(\text{KLD}_\text{global} = \tfrac12\,\hat{\boldsymbol\delta}^\top \boldsymbol\Omega^{-1} \hat{\boldsymbol\delta}\) is the location term of the VB shift (Section~\ref{sec-vb}).
The function \texttt{compare()} tabulates the log-evidences, log Bayes factors (relative to the best model), DIC, \(p_D\) (see below), and optionally any Bayesian fit indices, in a single summary table.
An example of comparing two nested models is given in Section~\ref{sec-circfa}.
Unlike bridge sampling \autocite{gronau2020bridgesampling}, which requires MCMC output and can be sensitive to the proposal distribution, the Laplace evidence is deterministic and available at no additional cost.
Like any Bayes factor, the result depends on the priors, and diffuse priors on the extra parameters of a larger model penalise it, so Bayes factors are best reported together with a check of their sensitivity to the priors \autocite{schad2023workflow,tendeiro2019review}.

\textbf{Deviance-based indices.} The Deviance Information Criterion \autocite[DIC;][]{spiegelhalter2002bayesian}, \(\text{DIC} = \bar D + p_D\) with \(p_D\) the effective number of parameters, and the Wishart-based posterior predictive \(p\)-value (PPP) of \textcite{levy2011bayesian} are computed from the posterior samples in the usual way.
For two-level models, the PPP follows \texttt{blavaan} in replicating data sets with the observed cluster design and scoring them against the saturated two-level model.

\textbf{Bayesian fit indices.} \texttt{bfit\_indices()} returns posterior distributions of BRMSEA, BGammaHat (and its adjusted form), BMc, BCFI, BTLI, and BNFI, computed from the posterior samples \autocite{garnier2020adapting,asparouhov2021advances,hoofs2018evaluating}.
The incremental indices are computed against an independence model that \texttt{INLAvaan} fits automatically.

\subsubsection{Standardised Solutions}\label{sec-defined}

\texttt{standardisedsolution()} applies the usual standardisation formulae \autocite{rosseel2012lavaan,kline2023principles} to each posterior draw of the raw parameters, propagating the full posterior uncertainty---including the covariance between loadings and residual variances---through the nonlinear transform.
The result is a data frame in \texttt{lavaan}'s familiar format augmented with posterior means, standard deviations, and credible intervals.

\begin{tcolorbox}[enhanced jigsaw, arc=.35mm, bottomrule=.15mm, bottomtitle=1mm, breakable, colback=white, colbacktitle=quarto-callout-note-color!10!white, colframe=quarto-callout-note-color-frame, coltitle=black, left=2mm, leftrule=.75mm, opacityback=0, opacitybacktitle=0.6, rightrule=.15mm, title=\textcolor{quarto-callout-note-color}{\faInfo}\hspace{0.5em}{Implication for Users}, titlerule=0mm, toprule=.15mm, toptitle=1mm]

A single fitted \texttt{INLAvaan} object supports a rich post-hoc workflow, which includes factor scores and Bayesian fit indices with full posterior uncertainty, Bayes factors for model comparison, and standardised solutions.
The computational cost of all post-hoc inference is dominated by the \texttt{nsamp} posterior draws from the copula, which at the default of 1,000 samples adds minor overhead for most models.

\end{tcolorbox}

\section{Bifactor Circumplex Factor Analysis}\label{sec-circfa}

\begin{figure}

\begin{minipage}{0.50\linewidth}

\pandocbounded{\includegraphics[keepaspectratio]{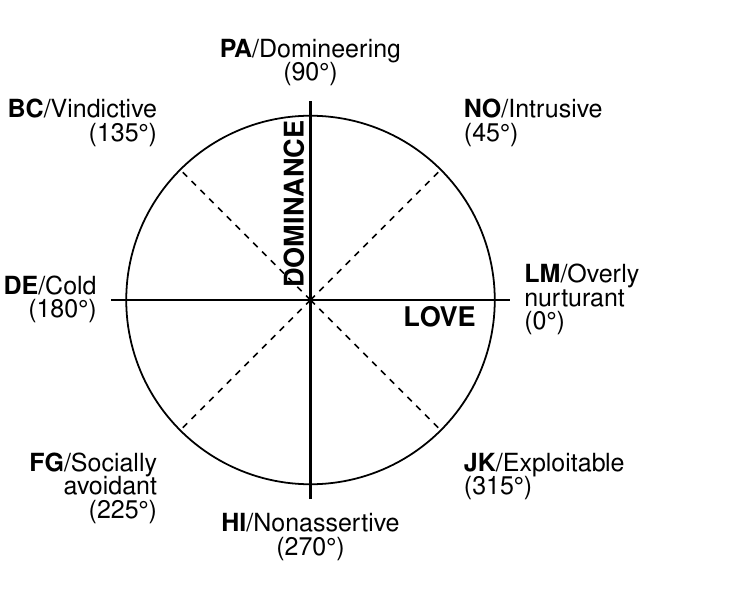}}

\end{minipage}%
\begin{minipage}{0.50\linewidth}

\fontsize{8.0pt}{10.0pt}\selectfont
\begin{tabular*}{\linewidth}{@{\extracolsep{\fill}}>{\raggedright\arraybackslash}p{\dimexpr 0.12\linewidth -2\tabcolsep-1.5\arrayrulewidth}>{\raggedright\arraybackslash}p{\dimexpr 0.88\linewidth -2\tabcolsep-1.5\arrayrulewidth}}
\toprule
Octant & Sample IIP item \\ 
\midrule\addlinespace[2.5pt]
{\bfseries PA} & I try to control other people too much. \\ 
{\bfseries BC} & It is hard to put someone else’s needs before my own. \\ 
{\bfseries DE} & It is hard to show affection to people. \\ 
{\bfseries FG} & I am too afraid of other people. \\ 
{\bfseries HI} & It is hard to be firm when I need to be. \\ 
{\bfseries JK} & It is hard to say ‘no’ to other people. \\ 
{\bfseries LM} & I put other people’s needs before my own too much. \\ 
{\bfseries NO} & I open up to people too much. \\ 
\bottomrule
\end{tabular*}

\end{minipage}%

\caption{\label{fig-iip}The Inventory of Interpersonal Problems (IIP) circumplex. Left: the eight octants at their ideal equi-spaced angles on the interpersonal circle, with solid lines marking the Love (horizontal) and Dominance (vertical) axes. Right: a sample item measuring each octant on a 0--4 Likert scale.}

\end{figure}

This is the first of two illustrative examples applying \texttt{INLAvaan} to analyses where fast fitting pays off.
In interpersonal psychology, personality constructs are arranged on a continuous circular space defined by two orthogonal dimensions: \emph{Dominance} and \emph{Love}.
Each subject's interpersonal style is characterised by an angular position \(\varphi_s\) and a radius \(R_s\) (style distinctiveness), while test items are anchored at known angles \(\delta_i\) around the circumplex \autocite{gurtman1992construct,wiggins1979psychological}.
For example, the Inventory of Interpersonal Problems \autocite[IIP;][]{horowitz1988inventory} measures \(p = 64\) items across eight octants (eight items each), as depicted in Figure~\ref{fig-iip}.
Unlike a single linear scale, a circle captures the dual nature of human interaction, which is driven by simultaneous forces of dominance (agency, status, power, or control) and love (friendliness, communion, solidarity, or warmth).
This circular structure models the continuous ``push and pull'' of interpersonal theory, illustrating how specific behaviours naturally evoke predictable reactions from others, such as dominance inviting submission, or hostility breeding hostility \autocite{kiesler19831982}.
By mapping maladaptive behaviours onto this space, the IIP provides a powerful framework that reveals not just a subject's internal distress, but the rigid, predictable ways they disrupt their social environment.

Beyond angular position, each subject is prescribed a general \emph{elevation} \(\tau_s\)---a tendency to endorse items highly regardless of circular position (akin to an acquiescence factor)---producing the bifactor measurement model
\begin{equation}\protect\phantomsection\label{eq-circfa}{
y_{si} = \tau_s + a_i \, R_s \cos(\varphi_s - \delta_i) + \epsilon_{si}, \qquad \epsilon_{si} \sim \N(0, \theta_i),
}\end{equation}
where \(a_i > 0\) is an item-specific \emph{amplitude} for \(i = 1, \ldots, p\) and \(s = 1, \ldots, n\).
The cosine kernel implies that a subject responds most favourably to items near their own angle and least favourably to items on the opposite side.
Expanding via \(\cos(\alpha - \beta) = \cos\alpha\cos\beta + \sin\alpha\sin\beta\) and defining the two orthogonal dimensions \(\text{Lov}_s = R_s\cos\varphi_s\) and \(\text{Dom}_s = R_s\sin\varphi_s\) converts this from polar to Cartesian coordinates \autocite{fabrigar1997conceptual}.
The result is a standard orthogonal CFA \autocite{wilson2013confirmatory} that any SEM software can estimate:
\begin{equation}\protect\phantomsection\label{eq-circfa-linear}{
%\underbrace{\begin{pmatrix} y_{s1} \\ y_{s2} \\ \vdots \\ y_{sp} \end{pmatrix}}_{\boldsymbol{y}_s}
\mathbf y_s
=
\underbrace{\begin{pmatrix}
\lambda_1^{(\tau)} & \lambda_1^{(\text{L})} & \lambda_1^{(\text{D})} \\
\lambda_2^{(\tau)} & \lambda_2^{(\text{L})} & \lambda_2^{(\text{D})} \\
\vdots & \vdots & \vdots \\
\lambda_p^{(\tau)} & \lambda_p^{(\text{L})} & \lambda_p^{(\text{D})}
\end{pmatrix}}_{\boldsymbol{\Lambda}}
\underbrace{\begin{pmatrix} \tau_s \\ \text{Lov}_s \\ \text{Dom}_s \end{pmatrix}}_{\boldsymbol{\eta}_s}
+ \ \boldsymbol{\epsilon}_s \, ,
}\end{equation}
with \(\boldsymbol{\eta}_s \sim \N(\boldsymbol{0},\, \operatorname{diag}(\psi_\tau, \psi_{\text{L}}, \psi_{\text{D}}))\), \(\boldsymbol{\epsilon}_s \sim \N(\boldsymbol{0},\, \operatorname{diag}(\theta_1, \ldots, \theta_p))\), and \emph{circular loadings} that satisfy \(\lambda_i^{(\text{L})} = a_i\cos\delta_i\) and \(\lambda_i^{(\text{D})} = a_i\sin\delta_i\).
The elevation loadings \(\lambda_i^{(\tau)}\) generalise the implicit unit coefficients on \(\tau_s\) in (\ref{eq-circfa}).

In frequentist ML estimation, all circular loadings are typically fixed at their theoretical values (\(\cos\delta_i\), \(\sin\delta_i\)) and the general-factor loadings at \(\lambda_i^{(\tau)} = 1\), because freeing them without constraints leads to rotational indeterminacy and nonsensical solutions \autocite{wilson2013confirmatory}.
The equal-variance constraint \(\psi_{\text{L}} = \psi_{\text{D}}\) is also imposed to preserve circular symmetry \autocite{browne1992circumplex}.
While tractable, this approach is too rigid when item angles deviate from ideal spacing---as they invariably do in practice \autocite{weide2021bayesian}.

A Bayesian alternative replaces these hard constraints with informative priors centred at the theoretical values (e.g., \(\lambda_i^{(\text{L})} \sim \N(\cos\delta_i,\, \sigma^2)\) with small \(\sigma\)), letting data pull the posterior towards the true item positions while retaining the confirmatory structure.
As loadings are free to deviate, the analyst can directly assess how well the circumplex holds, a diagnostic unavailable under fixed-loading approaches.
These same priors regularise the resulting 256-parameter space and guard against the Heywood cases and inadmissible solutions that ML routinely encounters at this dimensionality \autocite{chen2001improper}.
The rotation is softly identified by the collective mass of all informative priors rather than by arbitrary marker-item choices; detailed specifications are given in the Model Fit section below.

This is a setting in which \texttt{INLAvaan} excels, addressing three barriers that have made Bayesian circumplex modelling laborious.
First, \textbf{computational speed}: fitting the 256-parameter soft-constraint model takes about 12 seconds, so iterative model building with many refits remains practical.
Second, \textbf{syntactic accessibility}: the \texttt{prior()} modifier allows each loading to receive its own informative prior directly inside the \texttt{lavaan} formula, eliminating the need for convoluted external parameter tables or bespoke Stan code that would otherwise be required to attach 128 distinct priors to specific loadings.
Third, \textbf{seamless uncertainty propagation}: once fitted, \texttt{sampling()} draws from the joint posterior of all parameters efficiently, so quantities that are nonlinear functions of the loadings, such as octant angles \(\hat\delta_i\) and amplitudes \(\bar a_i\), inherit full posterior uncertainty.
The \texttt{predict()} method extends this to latent factor scores, returning posterior draws of individual-level angular positions and radii.

\subsection{Data Simulation}\label{data-simulation}

Because the IIP instrument and its clinical data sets are strictly proprietary, empirical evaluation using raw response data is precluded.
Indeed, modelling of the circumplex bifactor structure has historically relied almost exclusively on aggregated data---specifically, reducing the 64 items to eight unit-weighted sum scores (or octant scores) to avoid computational bottlenecks, analysing them via simplified factor models or the Structural Summary Method \autocite[SSM;][]{gurtman1992construct}.
We instead simulate data from the circumplex measurement model to demonstrate the workflow, which allows us to evaluate parameter recovery against known ground truth.
The simulated data set mirrors the IIP structure \autocite{horowitz2000ip64}: \(p = 64\) items (eight per octant) on a five-point Likert scale (0--4), and a realistic clinical inventory sample size of \(n = 800\) respondents.
The design is challenging in four respects:

\begin{enumerate}
\def\labelenumi{\arabic{enumi}.}
\tightlist
\item
  \textbf{Moderate sample, many parameters.} With 256 free parameters, the parameter-to-observation ratio is demanding for any estimator.
\item
  \textbf{Non-equispaced angles.} Data-generating octant angles follow the empirical values of \textcite{weide2021bayesian} (e.g., PA at 103° instead of the ideal 90°), so the circumplex deviates from textbook spacing.
\item
  \textbf{Unequal amplitudes.} Octant-specific amplitudes range from 0.7 to 1.2, breaking the perfect-circle assumption.
\item
  \textbf{Heterogeneous noise.} Item residual standard deviations are drawn from \(\text{Uniform}(0.3, 1.2)\).
\end{enumerate}

\subsection{Model Fit}\label{model-fit}

This example runs one pass of the Bayesian workflow: we fit a model from theory and a revised model, compare the two, check the computation and the fit of both, and check how much the conclusions depend on the priors.
We fit two models using standard \texttt{lavaan} syntax, differing only in how the circular loadings are treated.
Since the model specification involves 64 items across three factors, each requiring its own loading and (in the soft-constraint case) its own individually centred prior, writing the syntax by hand would be tedious and error-prone.
We therefore construct it programmatically in a short loop (see Supplementary Material for the R code).

\textbf{Model 1 (Fixed loadings).} All general-factor loadings are fixed to 1, circular loadings are fixed at the ideal trigonometric values (\(\cos\delta_i^{\text{ideal}}\), \(\sin\delta_i^{\text{ideal}}\)), the two circular factor variances are equality-constrained, and all factors are orthogonal.
This mirrors the standard ML approach of \textcite{wilson2013confirmatory} and yields 66 free parameters.

\textbf{Model 2 (Soft constraints).} All 192 loadings (64 per factor) are freely estimated, with all three factor variances fixed to 1 and mutual orthogonality imposed.
Each circular loading receives an informative prior centred at its ideal trigonometric value with \(\sigma = 0.2\), e.g., \(\lambda_i^{(\text{D})} \sim \N(\sin\delta_i^{\text{ideal}},\, 0.2^2)\).
The general-factor loadings are given default diffuse priors.
No marker items are fixed: the rotation is \emph{softly identified} by the collective mass of all 128 informative circular-loading priors, eliminating the arbitrary marker-item choice that classical CFA requires and avoiding the systematic bias that a misplaced anchor can introduce.
This yields 256 free parameters.
The key syntactic feature is the \texttt{prior()} modifier, which allows each loading to receive its own prior within the \texttt{lavaan} formula:

\begin{Shaded}
\begin{Highlighting}[]
\CommentTok{\# Excerpt: specifying per{-}loading priors for the Dominance factor}
\NormalTok{Dominance }\OtherTok{=}\ErrorTok{\textasciitilde{}} \FunctionTok{prior}\NormalTok{(}\StringTok{\textquotesingle{}normal(1, 0.2)\textquotesingle{}}\NormalTok{)}\SpecialCharTok{*}\NormalTok{PA\_1 }\SpecialCharTok{+}           \CommentTok{\# sin(90°)}
  \FunctionTok{prior}\NormalTok{(}\StringTok{\textquotesingle{}normal(0.7071, 0.2)\textquotesingle{}}\NormalTok{)}\SpecialCharTok{*}\NormalTok{NO\_1 }\SpecialCharTok{+}                 \CommentTok{\# sin(45°)}
  \FunctionTok{prior}\NormalTok{(}\StringTok{\textquotesingle{}normal(0, 0.2)\textquotesingle{}}\NormalTok{)}\SpecialCharTok{*}\NormalTok{LM\_1 }\SpecialCharTok{+}\NormalTok{ ...                  }\CommentTok{\# sin(0°)}
\end{Highlighting}
\end{Shaded}

The priors encode the circumplex geometry \emph{softly}: opposite octants receive negated prior centres (e.g., PA items near \(\sin 90^\circ = 1\) and HI items near \(\sin 270^\circ = -1\)), so departures from strict opposition can be discovered from the data.
Both models are fitted with \texttt{acfa()}.
Setting \texttt{std.lv\ =\ TRUE} fixes all factor variances to 1 and disables \texttt{lavaan}'s default reference indicator constraints, so that every loading is free to be governed by its prior:

\begin{Shaded}
\begin{Highlighting}[]
\NormalTok{R}\SpecialCharTok{\textgreater{}}\NormalTok{ fit\_fixed }\OtherTok{\textless{}{-}} \FunctionTok{acfa}\NormalTok{(}
\SpecialCharTok{+}    \AttributeTok{model =}\NormalTok{ mod\_fixed,}
\SpecialCharTok{+}    \AttributeTok{data =}\NormalTok{ iip\_data,}
\SpecialCharTok{+}    \AttributeTok{marginal\_correction =} \StringTok{"shortcut\_fd"}
\SpecialCharTok{+}\NormalTok{  )}
\NormalTok{R}\SpecialCharTok{\textgreater{}}\NormalTok{ fit\_soft }\OtherTok{\textless{}{-}} \FunctionTok{acfa}\NormalTok{(}
\SpecialCharTok{+}    \AttributeTok{model =}\NormalTok{ mod\_soft,}
\SpecialCharTok{+}    \AttributeTok{data =}\NormalTok{ iip\_data,}
\SpecialCharTok{+}    \AttributeTok{std.lv =} \ConstantTok{TRUE}\NormalTok{,}
\SpecialCharTok{+}    \AttributeTok{marginal\_correction =} \StringTok{"shortcut\_fd"}
\SpecialCharTok{+}\NormalTok{  )}
\end{Highlighting}
\end{Shaded}

\begin{Shaded}
\begin{Highlighting}[]
\NormalTok{R}\SpecialCharTok{\textgreater{}} \FunctionTok{print}\NormalTok{(fit\_soft)}
\end{Highlighting}
\end{Shaded}

\begin{verbatim}
INLAvaan 0.3.3 ended normally after 771 iterations

  Estimator                                      BAYES
  Optimization method                           NLMINB
  Number of model parameters                       256

  Number of observations                           800

Model Test (User Model):

   Marginal log-likelihood                  -58725.032 
   PPP (Chi-square)                              0.100 
\end{verbatim}

Fitting took about 6 seconds and 12 seconds for the fixed and soft models, respectively, using \texttt{marginal\_correction\ =\ "shortcut\_fd"} for both because of the size of the soft model (Section~\ref{sec-volume-correction}).
Bayesian model comparison via \texttt{compare()} strongly favours the soft-constraint model, with a log Bayes factor of 1,477 and a drop in DIC of about 3,900 units.
The fit indices agree.
BGammaHat and BCFI, the indices recommended by \textcite{garnier2020adapting}, are 0.857 and 0.923 for the fixed model against 0.994 and 0.997 for the soft model, so only the soft model clears the conventional 0.95 benchmark; BRMSEA (0.051 against 0.010) points the same way.
Other fit measures (e.g., \texttt{"BTLI"}, \texttt{"BMc"}, \texttt{"BNFI"}) can be requested through the \texttt{fit.measures} argument of \texttt{compare()}.

\begin{Shaded}
\begin{Highlighting}[]
\NormalTok{R}\SpecialCharTok{\textgreater{}} \FunctionTok{compare}\NormalTok{(fit\_fixed, fit\_soft, }\AttributeTok{fit.measures =} \FunctionTok{c}\NormalTok{(}\StringTok{"BGammaHat"}\NormalTok{, }\StringTok{"BCFI"}\NormalTok{, }\StringTok{"BRMSEA"}\NormalTok{))}
\end{Highlighting}
\end{Shaded}

\begin{verbatim}
Bayesian Model Comparison (INLAvaan)
Models ordered by marginal log-likelihood

     Model npar Marg.Loglik     logBF    DIC      pD BRMSEA BGammaHat   BCFI
 fit_fixed   66   -60201.54 -1476.511 119991  66.484 0.0514    0.8574 0.9228
  fit_soft  256   -58725.03     0.000 116062 236.324 0.0101    0.9941 0.9972
\end{verbatim}

Neither model raises a warning in the computation checks of Section~\ref{sec-mode-hessian}, and the PPP rejects the fixed model (0.00) but not the soft model (0.10).
Refitting the soft model with \texttt{update()} under circular-loading prior standard deviations of 0.1 and 0.5 moves no octant angle by more than 2.6° and leaves the soft model strongly preferred (log Bayes factors of 1,476 and 1,390).
In all, the four fits (the two models and the two prior refits) took about 47 seconds, demonstrating that a full pass of the iterative Bayesian workflow is feasible with \texttt{INLAvaan} even for a 256-parameter model.

\subsection{Results}\label{results}

We examine parameter recovery and individual-level prediction for the soft-constraint model.

\subsubsection{Angle and Amplitude Recovery}\label{angle-and-amplitude-recovery}

In the soft-constraint approach, posterior loadings can be translated back into the circumplex geometry.
Following \textcite{weide2021bayesian}, we average the Dominance and Love loadings within each octant to obtain octant-level coordinates, from which the angle is recovered via \(\hat\delta_i = \mathrm{atan2}(\bar\lambda_i^{(\text{D})},\, \bar\lambda_i^{(\text{L})})\) and the amplitude via \(\bar{a}_i = \tfrac{1}{8}\sum_{j \in \mathcal{O}_i} \sqrt{(\hat\lambda_j^{(\text{D})})^2 + (\hat\lambda_j^{(\text{L})})^2}\), where \(\mathcal{O}_i\) indexes the eight items of octant \(i\).
To obtain posterior uncertainty on these derived quantities, we draw 2,000 parameter samples via \texttt{sampling()} and propagate each draw through the same pipeline:

\begin{Shaded}
\begin{Highlighting}[]
\NormalTok{R}\SpecialCharTok{\textgreater{}}\NormalTok{ samp }\OtherTok{\textless{}{-}} \FunctionTok{sampling}\NormalTok{(fit\_soft, }\AttributeTok{type =} \StringTok{"lavaan"}\NormalTok{, }\AttributeTok{nsamp =} \DecValTok{2000}\NormalTok{)}
\NormalTok{R}\SpecialCharTok{\textgreater{}} \FunctionTok{dim}\NormalTok{(samp)}
\end{Highlighting}
\end{Shaded}

\begin{verbatim}
[1] 2000  256
\end{verbatim}

\begin{figure}

\begin{minipage}{0.50\linewidth}

\pandocbounded{\includegraphics[keepaspectratio,alt={Octant angles}]{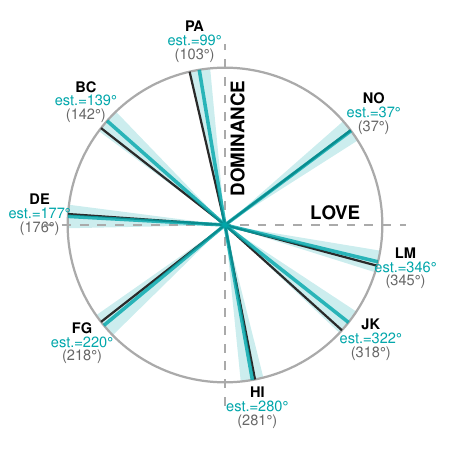}}

\subcaption{\label{}Octant angles}
\end{minipage}%
\begin{minipage}{0.50\linewidth}

\includegraphics[width=1\linewidth,height=\textheight,keepaspectratio,alt={Amplitude bias}]{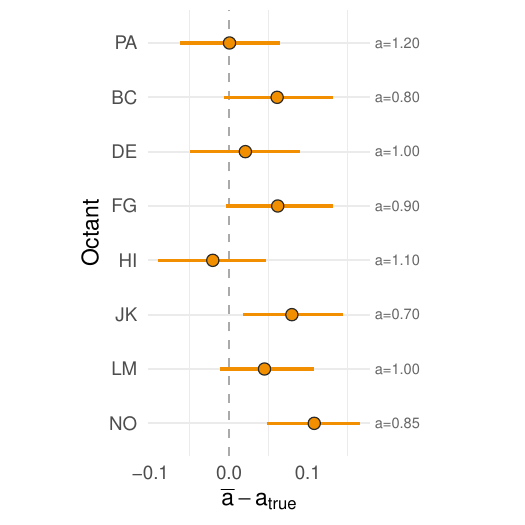}

\subcaption{\label{}Amplitude bias}
\end{minipage}%

\caption{\label{fig-circumplex-recovery}Recovered circumplex geometry from the soft-constraint model. Left: estimated octant angles (teal spokes) with 95\% credible wedges, overlaid on the true positions (black spokes). Right: bias in estimated mean octant amplitude \(\bar{a}_i\).}

\end{figure}

The function \texttt{sampling()} returns draws from the joint posterior of all SEM parameters.
Thus, any transformation of those draws yields a corresponding posterior sample, so summaries such as 95\% credible intervals for each octant's angular position and amplitude follow immediately, even for quantities not directly parameterised in the model.
Figure~\ref{fig-circumplex-recovery} (left) shows that the estimated angles closely track the true data-generating values, despite priors centred at the ideal equispaced positions; the opposition structure emerges naturally without explicit constraints.

The right panel in Figure~\ref{fig-circumplex-recovery} displays the amplitude bias \(\bar{a}_i - a_{i,\text{true}}\) with 95\% credible intervals.
All intervals are narrow and lie close to zero, so the model recovers the unequal octant amplitudes well.
The amplitudes of JK (Exploitable, \(a = 0.7\)) and NO (Intrusive, \(a = 0.85\)) are mildly overestimated, by about 0.08 and 0.11, though the angular positions of both remain well recovered.

\subsubsection{Factor Scores}\label{factor-scores}

Similar to \texttt{blavaan}, latent variable scores in \texttt{INLAvaan} are treated as a post-estimation procedure and must be requested separately.
For this 256-parameter model, \texttt{predict()} returns 2,000 draws in 1.8 seconds, each an \(800 \times 3\) matrix of (General, Dominance, Love) scores from which individual angular positions and radii are computed.

\begin{Shaded}
\begin{Highlighting}[]
\NormalTok{R}\SpecialCharTok{\textgreater{}}\NormalTok{ scores }\OtherTok{\textless{}{-}} \FunctionTok{predict}\NormalTok{(fit\_soft, }\AttributeTok{nsamp =} \DecValTok{2000}\NormalTok{)}
\NormalTok{R}\SpecialCharTok{\textgreater{}} \FunctionTok{print}\NormalTok{(scores)}
\end{Highlighting}
\end{Shaded}

\begin{verbatim}
Predicted values from inlavaan model (type = "lv")
Number of samples: 2000 
First sample:
   General Dominance    Love
1  -2.2177     0.415 -1.1648
2  -0.0412    -1.110  1.8783
3   0.4102     1.550  0.4556
4   0.2538    -0.674  1.0195
5   1.1637     0.245 -0.6271
6  -0.8404     0.350  0.5717
7   1.3215     1.418 -0.0685
8  -0.4060     1.429  0.5044
9   0.7638     0.695  0.0934
10  0.2360     1.693 -0.7143
# i 790 more rows
# i Use `summary()` to see summary statistics
\end{verbatim}

Angular positions and radii derived from these draws recover the true values well, except near the centre of the circumplex, where the angles are unstable (Figure~\ref{fig-polar-error} in Appendix B).

\section{Multilevel Mediation with Missing Data}\label{sec-safety}

Our next example for \texttt{INLAvaan} turns to a multilevel mediation model.
Few constructs in organisational psychology have attracted as much attention as \emph{psychological safety}, defined as the shared belief that a team is safe for interpersonal risk-taking \autocite{edmondson1999psychological}.
The construct was born from a paradox:
\textcite{edmondson1996learning} found that the best hospital nursing teams reported \emph{more} errors, not fewer, because they felt safe enough to speak up.
Expanding beyond clinical settings, Google's Project Aristotle identified psychological safety as the single strongest predictor of team effectiveness across 180 teams \autocite{duhigg2016what}, and now a large body of evidence even links it to learning behaviour, innovation, and error reporting in high-stakes environments ranging from aircraft carriers to operating theatres \autocite{edmondson2012teaming}.
Unpacking how leadership climate translates into this safety behaviour has become an inherently quantitative, mediational question, addressed through meta-analytic path models \autocite{frazier2017psychological} and multilevel mediation designs \autocite{neal2006study}.

\begin{figure}

\centering{

\includegraphics[width=1\linewidth,height=\textheight,keepaspectratio]{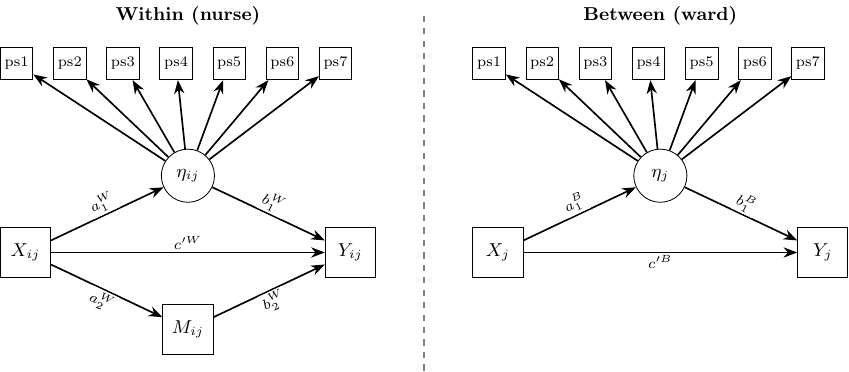}

}

\caption{\label{fig-safety-mediation}Multilevel mediation model. At the within-nurse level (left), Punitive Leadership (\(X\)) affects Error Concealment (\(Y\)) through two parallel pathways---Psychological Safety (\(\eta\), latent, measured by seven indicators) and Protocol Workarounds (\(M\), observed). At the between-ward level (right), only the psychological-safety pathway operates, since workarounds are modelled as a purely individual-level behaviour.}

\end{figure}

In healthcare, the stakes are existential.
When nurses perceive their supervisors as punitive, psychological safety erodes and staff stop speaking up about near-misses, work around safety protocols to avoid scrutiny, and in the worst case conceal medication errors entirely \autocite{edmondson1996learning,tucker2003why}.
The causal chain is a natural \emph{parallel mediation}, as seen in Figure~\ref{fig-safety-mediation}:
punitive leadership (\(X\)) simultaneously undermines the cognitive appraisal of safety (a latent construct, \(\eta\)) and drives observable behavioural shortcuts (\(M\)), both of which independently increase error concealment (\(Y\)).
Protocol workarounds, however, are a shift-by-shift behavioural response.
They vary meaningfully across individual nurses, but their ward-level mean carries no clear structural interpretation as a mediating \emph{mechanism}.
The shared psychological safety climate, by contrast, is genuinely a ward-level construct.
The between-ward model therefore retains only the psychological-safety pathway, while the within-ward model includes both mediators.

Measuring error concealment via self-report compounds the problem.
Nurses who are actively hiding errors, particularly those under the most punitive supervisors, are the most likely to skip the concealment question entirely \autocite{tourangeau2007sensitive}.
Dropping these incomplete cases systematically deletes the exact variance the model is trying to explain, severely biasing the structural parameters.
A credible analysis must therefore handle systematic missingness and, ideally, recover plausible individual-level scores for the nurses who refused to answer.

In these healthcare settings, researchers rely heavily on structural equation modelling to trace how unit-level leadership drives individual nurses' safety behaviours and error reporting \autocites[e.g.,][]{seo2022mediating,elhihi2025mediating}, yet accurately estimating 1-1-1 \textbf{multilevel mediation} paths with a \textbf{latent mediator}, parallel observed and latent indirect effects, and \textbf{systematic missingness} on a sensitive outcome remains a computational challenge.
This example pushes the envelope of what a single fitting call can achieve.
\texttt{INLAvaan} bridges this gap, delivering both the structural estimates and the handling of individual-level missingness in seconds.

\subsection{Data Simulation}\label{data-simulation-1}

While the data are fully simulated due to the lack of a true ward-level latent mediator in available public data sets, every structural choice is grounded in the published empirical literature, making this a realistic stress-test of the methodology.
We place 450 ICU nurses in \(J = 30\) hospital wards (\(n_j = 15\) per ward)---a sample frame typical of single-hospital nursing studies---with parameter values, instrument items, and ward-level variance components calibrated to the effect sizes reported in Edmondson's \autocite*{edmondson1996learning,edmondson1999psychological} foundational work on unit-level psychological safety and Tucker and Edmondson's \autocite*{tucker2003why} research on protocol workarounds.
The data-generating process involves four variables:

\begin{itemize}
\tightlist
\item
  \(X_{ij}\): \textbf{Punitive Leadership.} Each nurse's rating of their ward supervisor's hostility (continuous, centred).
\item
  \(\eta_{ij}\): \textbf{Psychological Safety} (latent). The nurse's inner appraisal of ward climate, measured by all seven items of \textcite[382]{edmondson1999psychological}:

  \begin{enumerate}
  \def\labelenumi{\arabic{enumi}.}
  \tightlist
  \item
    \emph{ps1} -- If you make a mistake on this team, it is often held against you.
  \item
    \emph{ps2} -- Members of this team are able to bring up problems and tough issues.
  \item
    \emph{ps3} -- People on this team sometimes reject others for being different.
  \item
    \emph{ps4} -- It is safe to take a risk on this team.
  \item
    \emph{ps5} -- It is difficult to ask other members of this team for help.
  \item
    \emph{ps6} -- No one on this team would deliberately act in a way that undermines my efforts.
  \item
    \emph{ps7} -- Working with members of this team, my unique skills and talents are valued and utilized.
  \end{enumerate}

  Factor loadings and uniquenesses are extracted from the published inter-item correlation matrix via standardised one-factor analysis.
\item
  \(M_{ij}\): \textbf{Protocol Workarounds.} The number of times the nurse bypassed a safety protocol this month (observed, continuous).
\item
  \(Y_{ij}\): \textbf{Medication Error Concealment.} A self-report scale measuring how actively the nurse covered up or failed to report a dosage error (observed, continuous).
\end{itemize}

\begin{longtable}[]{@{}
  >{\raggedright\arraybackslash}p{(\linewidth - 6\tabcolsep) * \real{0.1220}}
  >{\centering\arraybackslash}p{(\linewidth - 6\tabcolsep) * \real{0.1707}}
  >{\centering\arraybackslash}p{(\linewidth - 6\tabcolsep) * \real{0.1707}}
  >{\raggedright\arraybackslash}p{(\linewidth - 6\tabcolsep) * \real{0.5366}}@{}}
\caption{Data-generating structural coefficients for the multilevel mediation model.}\label{tbl-safety-params}\tabularnewline
\toprule\noalign{}
\begin{minipage}[b]{\linewidth}\raggedright
Path
\end{minipage} & \begin{minipage}[b]{\linewidth}\centering
Within
\end{minipage} & \begin{minipage}[b]{\linewidth}\centering
Between
\end{minipage} & \begin{minipage}[b]{\linewidth}\raggedright
Meaning
\end{minipage} \\
\midrule\noalign{}
\endfirsthead
\toprule\noalign{}
\begin{minipage}[b]{\linewidth}\raggedright
Path
\end{minipage} & \begin{minipage}[b]{\linewidth}\centering
Within
\end{minipage} & \begin{minipage}[b]{\linewidth}\centering
Between
\end{minipage} & \begin{minipage}[b]{\linewidth}\raggedright
Meaning
\end{minipage} \\
\midrule\noalign{}
\endhead
\bottomrule\noalign{}
\endlastfoot
\(X \to \eta\) & \(a_1 = -0.45\) & \(a_1 = -0.45\) & Punitive leadership erodes psychological safety \\
\(X \to M\) & \(a_2 = 0.35\) & & Punitive leadership drives workarounds \\
\(\eta \to Y\) & \(b_1 = -0.40\) & \(b_1 = -0.40\) & Low psychological safety increases concealment \\
\(M \to Y\) & \(b_2 = 0.30\) & & More workarounds increase concealment \\
\(X \to Y\) & \(c' = 0.12\) & \(c' = 0.12\) & Direct effect on concealment \\
\end{longtable}

The data-generating structural coefficients are given in Table~\ref{tbl-safety-params}.
At the between-ward level, the sole indirect pathway through psychological safety is \(a_1 b_1 = 0.180\) and the total effect is \(0.300\).
Within wards, both indirect channels operate: \(a_1 b_1 = 0.180\) (safety climate) and \(a_2 b_2 = 0.105\) (workarounds), giving a total within-level effect of \(0.405\).
Ward-level random intercepts for all variables (SDs ranging from 0.15 to 0.50) induce the clustered dependence.

To mirror the systematic dropout that hospital administrators actually encounter, we deliberately impose MAR missingness on \(Y\) via a logistic propensity model.
We assign the highest dropout probability to nurses under the most punitive supervisors who also perform the most workarounds, leaving administrators with a data set in which the wards they most need to monitor are also the most incomplete.
Since dropout depends only on the fully observed \(X\) and \(M\), not on the missing \(Y\) itself, the missingness mechanism is ignorable and the MAR assumption holds.
The procedure yields approximately 15.6\% missing \(Y\) values.

\subsection{Model Specification and Fit}\label{model-specification-and-fit}

The model uses the `\texttt{level:}' syntax from \texttt{lavaan} to define the within-ward and between-ward submodels.
The confirmatory factor analysis for Psychological Safety (\(\eta\), coded as \texttt{PS}) appears at both levels with all seven indicators, the first loading fixed to 1 for identification.
All structural paths are labelled, and the between-level indirect and total effects are defined via `\texttt{:=}':

\begin{Shaded}
\begin{Highlighting}[]
\NormalTok{mod\_safety }\OtherTok{\textless{}{-}} \StringTok{"}
\StringTok{  level: 1}
\StringTok{    PS =\textasciitilde{} ps1 + ps2 + ps3 + ps4 + ps5 + ps6 + ps7}
\StringTok{    PS \textasciitilde{}  aw1*X}
\StringTok{    M  \textasciitilde{}  aw2*X}
\StringTok{    Y  \textasciitilde{}  bw1*PS + bw2*M + cpw*X}

\StringTok{  level: 2}
\StringTok{    PS =\textasciitilde{} ps1 + ps2 + ps3 + ps4 + ps5 + ps6 + ps7}
\StringTok{    PS \textasciitilde{}  ab1*X}
\StringTok{    Y  \textasciitilde{}  bb1*PS + cpb*X}

\StringTok{  \# Between{-}level indirect and total effects}
\StringTok{  ind\_PS\_b := ab1 * bb1}
\StringTok{  total\_b  := cpb + ab1 * bb1}
\StringTok{"}
\end{Highlighting}
\end{Shaded}

Despite the complexity, \texttt{asem()} fits the model in about 6 seconds (with \texttt{cores\ =\ 14}, \texttt{nsamp\ =\ 10000}, and \texttt{test\ =\ "none"}).
The \texttt{cluster\ =\ "ward"} argument identifies the grouping column in \texttt{ward\_data}, and \texttt{fixed.x\ =\ FALSE} ensures that \(X\)'s variance is estimated freely at both levels (necessary in a 1-1-1 design where the predictor itself has within-ward and between-ward components).
The \texttt{missing\ =\ "ML"} argument (borrowing \texttt{lavaan}'s syntax) activates the casewise observed-data likelihood, so that each subject contributes to inference based on whichever variables they completed (the Bayesian analogue of full-information ML).
This must be specified explicitly, as the default is listwise deletion.

\begin{Shaded}
\begin{Highlighting}[]
\NormalTok{R}\SpecialCharTok{\textgreater{}}\NormalTok{ fit\_safety }\OtherTok{\textless{}{-}} \FunctionTok{asem}\NormalTok{(}\AttributeTok{model =}\NormalTok{ mod\_safety, }\AttributeTok{data =}\NormalTok{ ward\_data, }\AttributeTok{cluster =} \StringTok{"ward"}\NormalTok{,}
\SpecialCharTok{+}                     \AttributeTok{missing =} \StringTok{"ML"}\NormalTok{, }\AttributeTok{fixed.x =} \ConstantTok{FALSE}\NormalTok{, }\AttributeTok{cores =} \DecValTok{14}\NormalTok{, }
\SpecialCharTok{+}                     \AttributeTok{nsamp =} \DecValTok{10000}\NormalTok{, }\AttributeTok{test =} \StringTok{"none"}\NormalTok{)}
\NormalTok{R}\SpecialCharTok{\textgreater{}} \FunctionTok{print}\NormalTok{(fit\_safety)}
\end{Highlighting}
\end{Shaded}

\begin{verbatim}
INLAvaan 0.3.3 ended normally after 172 iterations

  Estimator                                      BAYES
  Optimization method                           NLMINB
  Number of model parameters                        51

  Number of observations                           450
  Number of clusters [ward]                         30
  Number of missing patterns -- level 1              2

Model Test (User Model):

   Marginal log-likelihood                   -6132.630 
\end{verbatim}

The default output reports unstandardised posterior summaries.
For fully standardised estimates, \texttt{standardisedsolution()} draws posterior samples of the free parameters internally and evaluates \texttt{lavaan}'s standardisation formulae at each draw, yielding posterior means, standard deviations, and credible intervals on the standardised scale:

\begin{Shaded}
\begin{Highlighting}[]
\NormalTok{R}\SpecialCharTok{\textgreater{}}\NormalTok{ stdsol }\OtherTok{\textless{}{-}} \FunctionTok{standardisedsolution}\NormalTok{(fit\_safety)}
\NormalTok{R}\SpecialCharTok{\textgreater{}} \FunctionTok{head}\NormalTok{(stdsol)}
\end{Highlighting}
\end{Shaded}

\begin{verbatim}
  lhs op rhs label est.std    se ci.lower ci.upper
1  PS =~ ps1         0.838 0.017    0.805    0.870
2  PS =~ ps2         0.816 0.017    0.780    0.849
3  PS =~ ps3         0.735 0.024    0.692    0.780
4  PS =~ ps4         0.853 0.014    0.826    0.879
5  PS =~ ps5         0.790 0.022    0.744    0.827
6  PS =~ ps6         0.789 0.021    0.739    0.822
\end{verbatim}

\subsection{Results}\label{results-1}

We examine the between-level indirect effects, ward-level factor scores, and posterior predictive imputations in turn.

\subsubsection{Indirect Effects}\label{indirect-effects}

The key estimands in this subsection are the between-level indirect and total effects, which quantify how much of the ward-level punitive-leadership effect travels through the psychological-safety climate.
\texttt{INLAvaan} stores skew-normal approximations to every posterior marginal---including defined quantities such as indirect effects---during model fitting itself.
No additional sampling step is needed: calling \texttt{plot(fit\_safety)} displays the posterior densities for all parameters at once, or selectively via the \texttt{params} argument:

\begin{Shaded}
\begin{Highlighting}[]
\FunctionTok{plot}\NormalTok{(fit\_safety)                                          }\CommentTok{\# all parameters}
\FunctionTok{plot}\NormalTok{(fit\_safety, }\AttributeTok{params =} \FunctionTok{c}\NormalTok{(}\StringTok{"ind\_PS\_b"}\NormalTok{, }\StringTok{"total\_b"}\NormalTok{))       }\CommentTok{\# selected subset}
\end{Highlighting}
\end{Shaded}

\begin{figure}[H]

\centering{

\includegraphics[width=1\linewidth,height=\textheight,keepaspectratio]{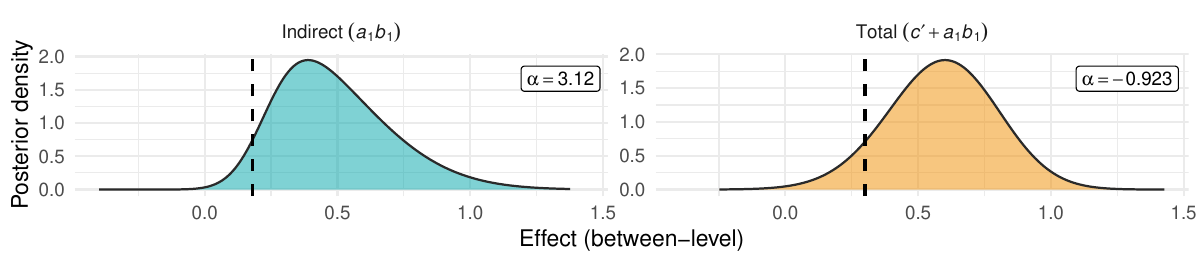}

}

\caption{\label{fig-indirect-density}Skew-normal posterior densities of the between-level effects. Left: indirect effect through psychological safety \(a_1 b_1\). Right: total effect \(c' + a_1 b_1\). Dashed lines mark the data-generating true values. Skew-normal shape parameter \(\alpha\) quantifies the asymmetry of the distribution, with larger absolute values indicating more skew.}

\end{figure}

The underlying density curves are also exported in the fitted object, making it straightforward to build custom visualisations.
Figure~\ref{fig-indirect-density} displays the between-level indirect and total effects with the data-generating truth overlaid.
The posterior for the indirect effect \(a_1 b_1\) through psychological safety is visibly right-skewed (\(\hat{\alpha} = 3.12\)), as expected for a product of two negative coefficients.
A symmetric credible interval from marginalising the joint Laplace (Gaussian) approximation would understate posterior mass in the upper tail and could misrepresent the strength of the mediation pathway.
Both the indirect and total effects place the bulk of their posterior mass above zero, consistent with a meaningful ward-level pathway through psychological safety.
With only \(J = 30\) wards, however, the credible intervals are wide, and appropriately so, since this is exactly the small-cluster regime where ML confidence intervals are known to undercover \autocite{mcneish2016usinga}.

\protect\phantomsection\label{cell-fig-indirect-density}

\subsubsection{Ward-Level Factor Scores}\label{ward-level-factor-scores}

Beyond the fixed structural paths, hospital administrators need to know \emph{which} wards are safest and which are most at risk.
In a multilevel model, each ward's position can be characterised by its between-level factor scores, the posterior distribution of the ward-specific random effects for psychological safety (PS), error concealment (\(Y\)), and punitive leadership (\(X\)).
\texttt{predict()} with \texttt{type\ =\ "lv"} and \texttt{level\ =\ 2L} draws these scores from the posterior, returning a list of matrices (one per sample, each \(J \times 3\)) that propagate both parameter uncertainty and shrinkage towards the grand mean:

\begin{Shaded}
\begin{Highlighting}[]
\NormalTok{R}\SpecialCharTok{\textgreater{}}\NormalTok{ ward\_lv }\OtherTok{\textless{}{-}} \FunctionTok{predict}\NormalTok{(fit\_safety, }\AttributeTok{type =} \StringTok{"lv"}\NormalTok{, }\AttributeTok{level =} \DecValTok{2}\NormalTok{L, }\AttributeTok{nsamp =} \DecValTok{2000}\NormalTok{)}
\NormalTok{R}\SpecialCharTok{\textgreater{}} \FunctionTok{length}\NormalTok{(ward\_lv)}
\end{Highlighting}
\end{Shaded}

\begin{verbatim}
[1] 2000
\end{verbatim}

\begin{Shaded}
\begin{Highlighting}[]
\NormalTok{R}\SpecialCharTok{\textgreater{}} \FunctionTok{str}\NormalTok{(ward\_lv[[}\DecValTok{1}\NormalTok{]])}
\end{Highlighting}
\end{Shaded}

\begin{verbatim}
 num [1:30, 1:3] -1.012 -0.25 0.612 0.098 0.762 ...
 - attr(*, "dimnames")=List of 2
  ..$ : NULL
  ..$ : chr [1:3] "PS" "Y" "X"
\end{verbatim}

\begin{Shaded}
\begin{Highlighting}[]
\NormalTok{R}\SpecialCharTok{\textgreater{}} \FunctionTok{summary}\NormalTok{(ward\_lv)}
\end{Highlighting}
\end{Shaded}

\begin{verbatim}
Mean of predicted values from inlavaan model

       PS       Y       X
1  -0.565  0.6877  0.3813
2  -0.175  0.1909  0.1132
3   0.393 -0.3410 -0.6297
4   0.309 -0.1802 -0.1904
5   0.337 -0.5382 -0.6843
6  -0.444  0.2081  0.2654
7   0.279  0.0485 -0.3911
8  -0.325  0.3901  0.5995
9  -0.251  0.1029 -0.3667
10 -0.429  0.5257  0.0479
# i 20 more rows
\end{verbatim}

\begin{figure}[htbp]

\centering{

\includegraphics[width=1\linewidth,height=\textheight,keepaspectratio]{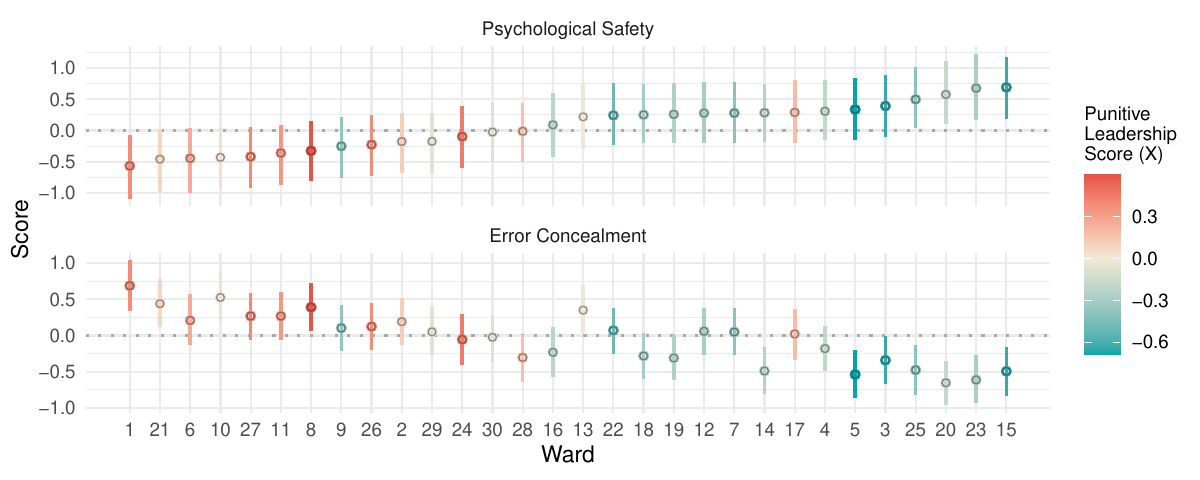}

}

\caption{\label{fig-ward-profiles}Ward-level factor scores ranked by posterior mean psychological safety. Top: Psychological Safety (PS). Bottom: Error Concealment (\(Y\)). Vertical bars are 95\% credible intervals; colour indicates the ward's punitive leadership score. Ward numbers on the horizontal axis correspond to the cluster identifiers in the data.}

\end{figure}

Figure~\ref{fig-ward-profiles} ranks the 30 wards by their posterior mean psychological safety, with 95\% credible intervals, alongside the corresponding ward-level error concealment.
Points are coloured by the ward's mean punitive leadership score: red wards have the most punitive supervisors, teal the least.
The mirror-image pattern captures the between-level mediation visually.
Wards at the bottom of the top panel (lowest psychological safety) tend to appear at the top of the bottom panel (highest concealment), and the colour gradient confirms that punitive leadership is the upstream driver.
A quality-improvement team could read this plot directly as a prioritised intervention list.

\protect\phantomsection\label{cell-fig-ward-profiles}

\subsubsection{Observed Value Imputation}\label{observed-value-imputation}

Beyond the structural parameters themselves, a hospital quality-improvement team needs individual-level predictions:
which nurses on which wards are most likely concealing errors, and with what uncertainty?
As described in Section~\ref{sec-predict}, \texttt{predict(fit,\ type\ =\ "ymis")} draws from the posterior predictive distribution of the missing outcomes at each of 2,000 posterior samples, drawing each ward's missing values jointly with its between-level values to propagate both parameter uncertainty and residual variability.
The output is a list of completed data frames that mirror the original data structure, with observed values left intact and only the missing cells filled by posterior draws.
Setting \texttt{ymis\_only\ =\ TRUE} returns just the imputed values as a named vector per sample (following the \texttt{blavaan::blavPredict()} convention), avoiding the overhead of reconstructing the full data frame.

\begin{Shaded}
\begin{Highlighting}[]
\NormalTok{R}\SpecialCharTok{\textgreater{}}\NormalTok{ imputed }\OtherTok{\textless{}{-}} \FunctionTok{predict}\NormalTok{(fit\_safety, }\AttributeTok{type =} \StringTok{"ymis"}\NormalTok{, }\AttributeTok{nsamp =} \DecValTok{2000}\NormalTok{)}
\NormalTok{R}\SpecialCharTok{\textgreater{}} \FunctionTok{print}\NormalTok{(imputed)}
\end{Highlighting}
\end{Shaded}

\begin{verbatim}
Predicted values from inlavaan model (type = "ymis")
Number of samples: 2000 
First sample:
      ps1    ps2     ps3    ps4    ps5     ps6     ps7       M      Y       X
1  -1.164  0.128 -1.5033 -2.001 -1.559  1.7126 -0.0735 -0.5416  0.587  0.3617
2  -0.193  1.547  1.0027 -0.568 -1.414 -0.6264  0.0385 -0.0405  0.566 -0.7343
3  -2.125 -1.376 -1.3734 -2.014 -0.699 -2.1463 -1.7316 -0.1659  1.839  0.1909
4  -2.929 -1.743 -2.0541 -2.091 -1.131  0.0596 -0.9871  2.1544  1.915  2.2400
5  -2.375 -0.626 -1.3964 -2.212 -0.987 -1.2145 -2.3414 -0.6157  0.953  1.0165
6  -2.492 -1.628 -2.1265 -1.783 -2.073 -1.5199 -2.4476  2.1168  1.905  1.2587
7   1.707  0.692  0.2260  0.557  1.468  1.4218  1.1448 -1.1731 -1.201  0.0454
8  -0.246 -1.440 -0.0601  0.490  0.509 -0.2900 -0.3279  0.3775  0.374 -1.0111
9  -1.706 -0.851 -2.1132 -1.294 -1.491 -1.7740 -0.0981  0.0310  0.620  0.8823
10 -1.538 -1.337 -2.5928 -0.808  0.733 -2.4031 -0.3819  0.0758  1.236  1.3595
# i 440 more rows
# i Use `summary()` to see summary statistics
\end{verbatim}

\begin{figure}[htbp]

\centering{

\includegraphics[width=1\linewidth,height=\textheight,keepaspectratio]{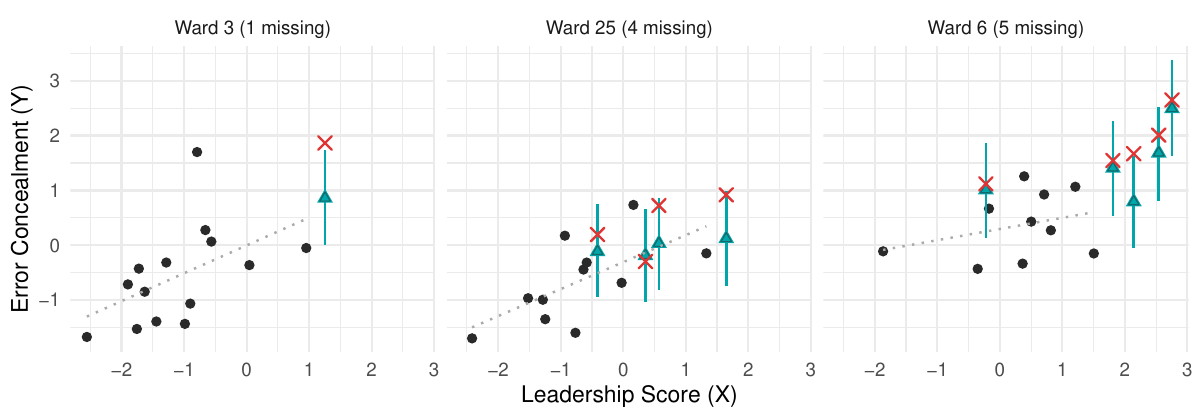}

}

\caption{\label{fig-imputation}Observed value imputation of Error Concealment for three hospital wards with low (Ward 3), moderate (Ward 25), and high (Ward 6) dropout. Black circles are observed \(Y\); teal triangles with 90\% credible intervals are posterior predictive imputations; and red crosses are withheld true values. Dotted grey lines are the ordinary least squares regression of observed \(Y\) on \(X\) for visual reference.}

\end{figure}

Figure~\ref{fig-imputation} illustrates the result for three wards with varying levels of dropout.
Black circles are nurses who completed the concealment question; teal triangles with 90\% credible intervals are posterior predictive imputations for those who left it blank; red crosses mark the withheld true values.
The imputed nurses cluster at higher \(X\) (more punitive leadership), consistent with the MAR mechanism, and 8 of the 10 withheld values fall within their intervals, in line with the nominal coverage.

\protect\phantomsection\label{cell-fig-imputation}

Across all 70 nurses with missing concealment data, the 90\% credible intervals achieved 87.1\% empirical coverage of the withheld true values, close to the nominal level.
Standard full-information maximum likelihood in \texttt{lavaan} cannot produce such individual-level imputations, and while MCMC certainly can, it comes at greater computational cost.
\texttt{INLAvaan} delivers them as a one-line byproduct of the already-fitted model.

\section{Open Science and Reproducibility}\label{open-science-and-reproducibility}

\texttt{INLAvaan} can be installed from CRAN, and the results in this paper were produced with version 0.3.3.
The development version is hosted on GitHub at \url{https://github.com/haziqj/INLAvaan}.
The repository also serves as the package's primary documentation hub, hosting the full function reference and a growing collection of vignettes covering common use cases to help new users get started.
The package includes a comprehensive test suite, run by GitHub Actions on every commit across multiple R versions and operating systems.
It covers key pipeline components and end-to-end non-regression checks against stored reference values to verify that code changes preserve prior behaviour.

The empirical data set used in this article (Political Democracy) is publicly accessible within the \texttt{lavaan} package.
All analysis scripts, model specifications, and simulated data files needed to reproduce the results in Sections \ref{sec-circfa} and \ref{sec-safety} are deposited at \url{https://osf.io/arqmh}, with step-by-step instructions and documented session information.

\section{Conclusion and Future Directions}\label{conclusion-and-future-directions}

This paper has described \texttt{INLAvaan}, an R package that turns the deterministic Bayesian SEM approximation of \textcite{jamil2026approximate} into a practical analysis tool.
Building the INLA approximation from scratch (rather than wrapping around the well-established \texttt{R-INLA} software) allowed the entire pipeline to be architected around SEM's specific structure, avoiding the overhead and reformulation costs that a general-purpose interface would impose.
By delegating only model algebra and syntax parsing to \texttt{lavaan}, the package inherits the full breadth of continuous normal-theory SEM specifications---including multigroup, multilevel, and missing-data models---with the custom approximation stages layered on top.
The two examples show that the speed of \texttt{INLAvaan} matters most when a model must be fitted many times.
The 256-parameter bifactor circumplex model in Section~\ref{sec-circfa} and the multilevel mediation model with missing data in Section~\ref{sec-safety} both show \texttt{INLAvaan} delivering calibrated posterior summaries, factor scores, and missing-data imputations quickly enough to make the iterative Bayesian workflow feasible.

Most of the computational cost lies in repeated evaluations of the log-posterior and its gradient, which the marginal scans and volume corrections require for every parameter.
Moving these evaluations from R into compiled code could therefore speed up the whole pipeline substantially.
Automatic differentiation (AD) in our view would add less, since the gradient is already available in closed form.
Its main gain would be exact Hessians and volume slopes in place of finite differences, though greater value would emerge in a standalone implementation independent of \texttt{lavaan}, where AD would facilitate the addition of new model classes without the need to derive their gradients by hand.

Certain limitations are inherent to the method and would persist even in a mature implementation \autocite[Sec. 5]{jamil2026approximate}.
In practice, expect the approximation to be poor when parameters are weakly identified, such as a factor with low loadings or strongly collinear predictors, and expect it to understate the lower tail of a variance close to zero; in such cases, MCMC checks are encouraged.
Furthermore, the approach requires a marginal likelihood, and ideally its gradient, that is cheap to evaluate, which normal-theory SEM provides in closed form.
Models whose latent variables cannot be integrated out cheaply, as for most mixture or nonlinear latent structures, or that use non-smooth shrinkage priors such as the Bayesian lasso of \textcite{chen2021generalized}, therefore remain the domain of MCMC.

Two main software limitations remain on the near-term roadmap.
First, \texttt{INLAvaan} currently supports only continuous outcomes under the normal-theory likelihood.
Binary and ordinal indicators, which are ubiquitous in psychometric practice, cannot be accommodated by simply reusing \texttt{lavaan}'s built-in categorical estimators (weighted least squares, in its full and diagonal forms, and pairwise maximum likelihood), because these are limited-information methods that optimise bivariate or pairwise marginals rather than a full likelihood surface \autocite{joreskog1990new,katsikatsou2012pairwise}, making them unsuitable for the profiling and Laplace machinery that \texttt{INLAvaan} relies on.
Progress therefore requires an efficient approximation to the intractable multivariate probit likelihood.
Second, cross-validation (CV) is the subject of ongoing work.
We share the view of \textcite{vehtari2017practical} that pointwise predictive assessment is the natural basis for comparing models in the Bayesian workflow, which refit-free leave-one-out CV (LOO-CV), including leave-one-cluster-out CV for multilevel models, and the Widely Applicable Information Criterion \autocite[WAIC;][]{watanabe2010asymptotic} will support once available.
Until then, the DIC remains the default criterion, and Bayes factors, which depend strongly on the priors, are best read alongside a check of their sensitivity to them.

On a smaller scale, Bayesian \(R^2\) summaries \autocite{gelman2019rsquared} and additional prior families, such as log-\(t\)-normal priors \autocite{martins2013bayesian}, LKJ priors \autocite{lewandowski2009generating}, and perhaps even fully user-supplied expression priors, are natural additions.
The modular design of the augmented parameter table (Section~\ref{sec-param-space}) makes this straightforward to implement, as new prior families require only an unnormalised log-density and its gradient.
These additions, and any other feature requests or contributions, are welcomed via the package's GitHub repository, and we hope that \texttt{INLAvaan} proves a useful addition to the Bayesian SEM toolkit.

\section*{Acknowledgements}\label{acknowledgements}
\addcontentsline{toc}{section}{Acknowledgements}

We are grateful to three anonymous reviewers whose comments substantially improved the manuscript.
This publication is based upon work supported by the King Abdullah University of Science and Technology (KAUST) Research Funding Office under Award No.~URF/1/6921-01-01.

\section*{References}\label{references}
\addcontentsline{toc}{section}{References}

\printbibliography[heading=none]

\newpage{}

\appendix

\section{Efficiency Benchmark for Marginal Profiling Grid Density}\label{efficiency-benchmark-for-marginal-profiling-grid-density}

\begin{figure}[H]

\centering{

\includegraphics[width=1\linewidth,height=\textheight,keepaspectratio]{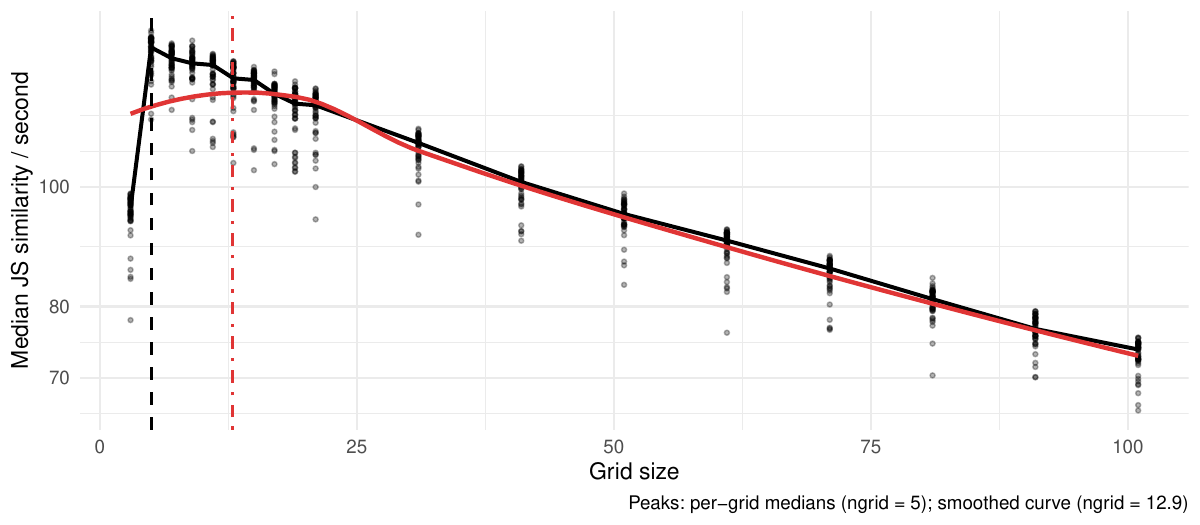}

}

\caption{\label{fig-effpertime}Efficiency (median Jensen--Shannon similarity to MCMC per second of fitting time) as a function of grid size (\(m = 31\) parameters). Each point is the median accuracy over all parameters divided by the fit time, for one of 50 repeated fits at that grid size; the black line connects the per-grid medians (peak at \texttt{ngrid} \(= 5\)) and the red curve is a smooth LOESS fit (peak at \texttt{ngrid} \(\approx 14\)). Efficiency declines beyond these points as the fitting cost grows faster than the accuracy.}

\end{figure}

We ran, for the Political Democracy example, 50 repeated benchmarks of various grid densities (\texttt{ngrid=3} to \texttt{101}) for the marginal profiling step.
In this benchmark, accuracy does not improve beyond about nine points (median Jensen--Shannon similarity 99.83\% at both 9 and 21 points), while the run time grows linearly with the number of grid points (by about 9\% from 9 to 21 points), so the efficiency in Figure~\ref{fig-effpertime} falls beyond its peak.

\protect\phantomsection\label{cell-fig-effpertime}

\section{Recovery of Individual Angular Positions}\label{recovery-of-individual-angular-positions}

\begin{figure}[H]

\centering{

\includegraphics[width=0.55\linewidth,height=\textheight,keepaspectratio]{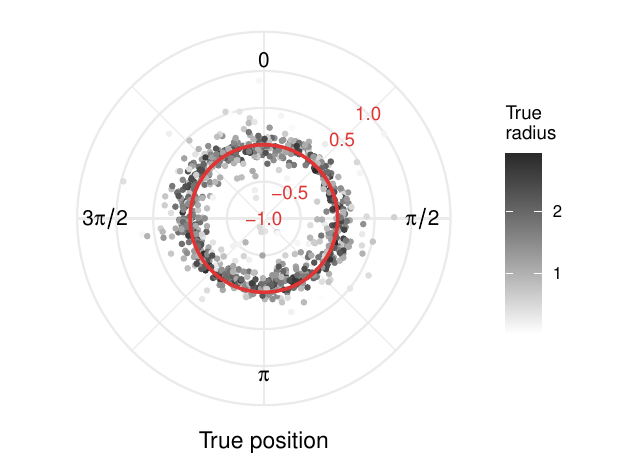}

}

\caption{\label{fig-polar-error}Polar projection of individual-level angular estimation error. The red circle at zero error represents perfect recovery. Points are coloured by true radius; angular estimates are inherently noisier for individuals near the centre.}

\end{figure}

Figure~\ref{fig-polar-error} displays the angular estimation error in polar coordinates for the circumplex example of Section~\ref{sec-circfa}.
Points near the red zero-error circle indicate accurate recovery, while colour encodes the true radius.
As expected, angular estimates are noisier for individuals near the origin of the circumplex, where \(\mathrm{atan2}\) is inherently unstable.
Overall, the soft-constraint model recovers individual-level angular positions and radii well, with the bulk of estimation error concentrated near the centre where the circumplex geometry is least informative.

\protect\phantomsection\label{cell-fig-polar-error}

\end{document}